\documentclass[10pt]{iopart}

\usepackage{iopams}

\expandafter\let\csname equation*\endcsname\relax
\expandafter\let\csname endequation*\endcsname\relax
\usepackage{physics}

\usepackage{amsfonts}
\usepackage{mathtools}
\usepackage{algorithm}
\usepackage{algpseudocode}
\usepackage{latexsym}

\usepackage{url}
\usepackage[dvipsnames]{xcolor}

\usepackage{mleftright}

\usepackage{graphicx}
\usepackage{caption}
\usepackage{subcaption}

\usepackage{siunitx}
\usepackage{tikz}
\usetikzlibrary{plotmarks, positioning} 
\usetikzlibrary{calc,intersections}
\usetikzlibrary{arrows, arrows.meta, positioning, shapes.geometric, shapes}
\usetikzlibrary{decorations.markings}
\usetikzlibrary{decorations.text}
\usetikzlibrary{matrix}

\usepackage{pgfplots}
\pgfplotsset{compat=newest}
\usepgfplotslibrary{groupplots,units}
\usepgfplotslibrary{colorbrewer}
\usepgfplotslibrary{colormaps}
\usepgfplotslibrary{fillbetween}

\usepackage{float}
\usepackage[section]{placeins} %
\usepackage{enumitem}
\usepackage{xifthen}
\usepgfplotslibrary{external}
\usepgfplotslibrary{fillbetween}
[
prefix={figures/}
]
\pgfplotsset{
	every tick label/.append style={font=\small},
	every axis/.append style={
		width = 0.45\textwidth,
		mark size = 2.85pt,
		axis line style = ultra thick,,
		legend style={cells={align=left}}, 
		legend style={draw=none},
		legend style={font=\small},
		legend style={fill=none, text opacity=1},
		colormap/jet
	},
    every axis plot/.append style={very thick} 
}

\usepackage{pdfpages}
\usepackage{booktabs}
\usepackage[toc,page]{appendix}
\usepackage{float}
\usepackage{slashbox}

\DeclarePairedDelimiterX{\mean}[1]{\langle}{\rangle}{#1}

\newcommand{\JointString}[2]{{#1#2}}

\newcommand{\Lorentz}{\mathcal{L}}
\newcommand{\MONKES}{{\texttt{MONKES}}}

\newcommand{\Eval}[1]
{
	#1|
}

\newcommand{\lambdac}{ \lambda_{\text{c}} }
\newcommand{\Bmax}{B_{\text{max}}}
\newcommand{\Bmin}{B_{\text{min}}}
\newcommand{\Dij}[1]{\widehat{D}_{#1}}
\newcommand{\epseff}{\epsilon_{\text{eff}}}
\newcommand{\qmarks}[1]{``#1''}
\newcommand{\GammaC}{\Gamma_{\text{c}}}
\newcommand{\GammaAlpha}{\Gamma_{\alpha}}
\newcommand{\Nfp}{N_{\text{fp}}}
\newcommand{\BpwO}{B_{\text{pwO}}}

\newcommand{\PlotDijCollisionality}[5]
{
	\begin{tikzpicture}
		\begin{axis}[    
			width=0.8\textwidth,    
			#1
			]
			
			\addplot+[#2] table [#3]{#4}; 
			#5
			
		\end{axis}
	\end{tikzpicture}
}

\newcommand{\FileLocation}{}
\newcommand{\PlotDijPiPow}[3]
{						 
			\foreach \F in {0.5, 0.6, 0.7, 0.8, 0.9, 1.0, 1.1, 1.2, 1.3, 1.4, 1.5, 1.6, 1.7, 1.8}		
			{				
				\renewcommand{\FileLocation}{
					\JointString{data/QI_pwQI_scan/wa\F}{pi_pow#3/monkes_Monoenergetic_Database.dat}
				}
			  \addplot[forget plot, #1] table [x expr=\F, #2]{\FileLocation};

			}

			\foreach \F in {0.9}		
			{				
				\renewcommand{\FileLocation}{
					\JointString{data/QI_pwQI_scan/wa\F}{pi_pow#3/monkes_Monoenergetic_Database.dat}
				}
				\addplot[#1] table [x expr=\F, #2]{\FileLocation};

			}
			
}

\begin{document}

\title[Evaluation of neoclassical transport in nearly quasi-isodynamic stellarator magnetic fields using MONKES]{Evaluation of neoclassical transport in nearly quasi-isodynamic stellarator magnetic fields using MONKES}

\author{F. J. Escoto, J. L. Velasco, I. Calvo and E. Sánchez}
\address{Laboratorio Nacional de Fusión, CIEMAT, 28040 Madrid, Spain}

\ead{fjavier.escoto@ciemat.es}
\vspace{10pt}
\begin{indented}
\item[] 
\end{indented}

\begin{abstract}    
Stellarator magnetic fields that are perfectly optimized for neoclassical transport (with levels of radial neoclassical transport comparable to tokamaks) are called omnigenous. Quasi-isodynamic magnetic fields are a subset of omnigenous magnetic fields in which the isolines of the magnetic field strength close poloidally, which grants them the additional property of producing zero bootstrap current. A frequent strategy in the quest for quasi-isodynamic configurations is to optimize the magnetic field indirectly by minimizing proxies that vanish in an exactly quasi-isodynamic field. The recently developed code {\MONKES} enables fast computations of the neoclassical radial transport and bootstrap current monoenergetic coefficients, and therefore facilitates enormously to assess the efficiency of such indirect approach. By evaluating the large database of intermediate configurations that led to the configuration CIEMAT-QI, the inefficiency of the indirect optimization strategy for minimizing the bootstrap current is illustrated. In addition, {\MONKES} is used to take the first steps in the exploration of a region of the configuration space of piecewise omnigenous fields, a novel family of optimized magnetic fields that has broadened the configuration space of stellarators with low levels of radial neoclassical transport.
\end{abstract}

%
\vspace{2pc}
\submitto{\NF}
%
%
\ioptwocol

\section{Introduction}\label{sec:Introduction}   

Stellarators are plasma confinement devices in which the magnetic field is entirely generated by external magnets. This is in contrast to tokamaks, which require a large toroidal current to generate the poloidal component of the magnetic field. The absence of this large toroidal current grants stellarators the benefit of avoiding current-induced instabilities and makes steady-state operation easier. Trapped particles drift preserving the second longitudinal adiabatic invariant $J$ \cite{dherbemont2022}. For a generic stellarator, the isosurfaces of $J$ are transversal to flux-surfaces, which implies that trapped particles quickly drift out of the device. Of course, this has a deleterious impact on the confinement which makes a generic stellarator invalid as a candidate for a future fusion reactor. 

Fortunately, there exists a class of stellarator magnetic fields for which collisionless trajectories are well confined in the absence of turbulence: \textit{omnigenous stellarators} \cite{CaryPhysRevLett,Cary1997OmnigenityAQ}. As in tokamaks, in an omnigenous stellarator the radial excursion that collisionless thermal particles experience vanishes when averaged along their orbits. This happens because in an omnigenous stellarator surfaces of constant $J$ and flux-surfaces coincide. Thus, in order to design stellarators, the magnetic field $\vb*{B}$ has to be carefully tailored so that the orbit-averaged radial drift is as small as possible, or equivalently, that $J$ approaches as much as possible a flux-function. This process is known as neoclassical stellarator optimization. Omnigenity \cite{CaryPhysRevLett,Cary1997OmnigenityAQ} imposes several restrictions to the isolines of the magnetic field strength $B:=|\vb*{B}|$ on a flux-surface. In an omnigenous stellarator, the isolines of $B$ must close poloidally, toroidally or helically around the torus. Omnigenous stellarators for which the isolines of $B$ close poloidally have the additional property of producing zero bootstrap current at low collisionality \cite{Helander_2009, Helander_2011_Bootstrap}. Magnetic fields of this kind are known as quasi-isodynamic (QI). The fact that QI stellarators have negligible bootstrap current makes them ideal candidates for a divertor design based on a particular structure of islands \cite{SunnPedersen_2019}. The effect of an uncontrolled bootstrap current on the magnetic configuration can alter significantly this structure and thus endanger the viability of this type of divertor. 


As in the optimization process a fast evaluation of neoclassical properties is required, stellarator optimization has been typically done \textit{indirectly}. For instance, one can minimize proxies which vanish in an exactly omnigenous configuration. An example of these quantities are ${\GammaC}$ \cite{NemovGammaC} and its recent refinement $\GammaAlpha$ \cite{Velasco_2021}, which measure how well the constant $J$ surfaces align with flux-surfaces. Another example is the \qmarks{effective ripple} $\epseff$ \cite{Nemov1999EvaluationO1}, which successfully encapsulates radial transport in the unfavourable $1/\nu$ regime. Minimizing $\epseff$ has the effect of shifting the $1/\nu$ regime to smaller values of the collisionality $\hat{\nu}$.  

In regard to parallel transport, there is no figure of merit that is sufficiently accurate for optimization purposes. Traditionally, QI stellarators have been neoclasically optimized keeping in mind the constraints to the topology of the isolines of $B$ established in \cite{CaryPhysRevLett,Cary1997OmnigenityAQ}. For instance, one could try to force the isolines of $B$ to close poloidally. Then, one \textit{trusts} that minimizing proxies for general omnigenity while ensuring that most of the isolines of $B$ close poloidally will minimize the bootstrap current. Remarkably, this strategy has proven to be successful in the past for designing QI stellarators with small levels of radial and parallel transport \cite{Sanchez_2023,Goodman_2023}. Despite this ultimate success, simply following this strategy has two main drawbacks. The first one is the imperfect correlation between proxies and the physical quantities that they represent, which may make the process inefficient. Additionally, this strategy precludes the possibility of finding non traditional optimized configurations. In other words, if there exist nearly omnigenous equilibria different from those defined in \cite{CaryPhysRevLett,Cary1997OmnigenityAQ}, they will hardly be found this way.

Direct optimization is important not only for obtaining better magnetic configurations but also for finding new ideal stellarator designs. In \cite{Bindel_2023}, fast ion confinement was improved by including guiding-center trajectories in the optimization loop. Naturally, configurations with very small levels of fast ion losses were produced. When inspecting the isolines of $B$ in magnetic coordinates of these configurations, the topology of constant $B$ contours within the flux-surface differed significantly from what would be expected of an omnigenous configuration. Therefore, it came as a surprise that some of these configurations displayed also small values of $\epseff$. Inspired by this result, a new family of optimized stellarators denominated \textit{piecewise omnigenous (pwO)} \cite{velasco2024piecewise}, has emerged. For these fields, the second adiabatic invariant $J$ is a flux-function only piecewisely, allowing jump discontinuities of $J$ on a flux-surface along the poloidal direction. Among other things, this implies that the topology of the isolines of $B$ in a pwO field is not as limited as for an omnigenous stellarator. Imposing $J$ to be constant in a particular region of the flux-surface constrains the isolines of $B$ in a similar way to the one presented in \cite{CaryPhysRevLett, Cary1997OmnigenityAQ} for omnigenous stellarators. Hence, as pwO magnetic fields have several regions in which $J$ is constant, the isolines are not necessarily forced to close poloidally, toroidally or helically. 

Modern tools and computation capabilities allow \textit{direct optimization} of neoclassical transport, opening new paths to explore the stellarator configuration space. Based on rigorously derived bounce-averaged equations \cite{dherbemont2022,Calvo_2017}, the fast neoclassical code {\texttt{KNOSOS}} \cite{KNOSOSJCP, KNOSOSJPP} made it possible to directly optimize radial transport in stellarators. More recent developments will help exploring configurations with small bootstrap current, which might not be extremely close to quasi-isodynamicity. Thanks to its speed of computation, the neoclassical code {\MONKES} \cite{escoto2024monkes} allows to address directly the optimization of the bootstrap current (and the optimization of neoclassical transport, in general) in stellarators. 

In this work we show two applications of {\MONKES}. In section \ref{sec:Correlations}, we present the results of a massive neoclassical transport evaluation of the large database of configurations generated during the optimization campaign that led to CIEMAT-QI4 \cite{Sanchez_2023}. This will shed light on how efficient the indirect approach to neoclassical stellarator optimization is. In spite of the fact that it allows to obtain optimized QI configurations, we will confirm that this strategy is inefficient for minimizing the bootstrap current. The second application is a first step in the exploration of a particular region of the stellarator configuration space with small levels of radial and parallel transport and which, computationally speaking, is very expensive to explore. We will use {\MONKES} to evaluate how radial transport and the bootstrap current evolve when a magnetic field moves from traditional quasi-isodynamicity to a pwO magnetic field which is close to quasi-isodynamicity.

\section{Evaluation of proxies for optimizing quasi-isodynamic configurations} \label{sec:Correlations}
In neoclassical optimization, one typically pursues omnigenous configurations by minimizing a \textit{cost function} $\chi$. One manner to express this function is as a distance
\begin{align}
	\chi^2 =
	\sum_{k}
	w_k^2 
	\left(
	\chi_k^{\text{target}}
	-
	\chi_k^{\text{eq}}
	\right)^2
	.
	\label{eq:Cost_function}
\end{align}
Here, $\chi_k$ stands for a specific proxy: a quantity that represents some property of the magnetic configuration. The value $\chi_k^{\text{target}}$ is the desired value for the aforementioned property and $\chi_k^{\text{eq}}$ is the actual value of $\chi_k$ for the magnetic configuration obtained by solving the magnetohydrodynamical equilibrium equation. For each value of $\chi$, the reciprocals of the scalars $w_k$ set an upper bound for the deviation  $|\chi_k^{\text{eq}} - \chi_k^{\text{target}}|\le |\chi/w_k|$. Hence, the weights $w_k$ set the relative importance of each proxy. Thus, a cost function is determined by a selection of proxies $\chi_k$, their target values $\chi_k^{\text{target}}$ and weights $w_k$.

The selection of the proxies $\{\chi_k\}$ is meant to parametrize the type of stellarator that one wishes to obtain. For instance, in neoclassical optimization, the proxies $\{\chi_k\}$ should represent as well as possible the neoclassical properties of the magnetic configuration while being fast to calculate. In order to reduce the value of $\chi^2$, several quantities of the magnetic configuration called \textit{variables} are modified by an optimizer (e.g. the modes of the Fourier representation of the last closed flux-surface).
  
Selecting a single cost function $\chi^2$ is usually insufficient for satisfying all the criteria required for the magnetic configuration. Therefore, an optimization campaign consists on successive optimization steps until the obtained magnetic field satisfies the given desiderata. Each optimization step is defined by a different cost function $\chi^2$. That is, from one step to another, the proxy selection $\{\chi_k\}$, their target values $\chi_k^{\text{target}}$ and/or their relative importance (i.e. the values $\{w_k\}$) are varied. How to successfully change the cost function from one step to the next is a non straightforward process which, in most cases, requires some experience, intuition and luck.

In order to neoclassically optimize QI configurations, the goal is to reduce not only $\Dij{11}$ but also $\Dij{31}$ as much as possible. Here, $\Dij{11}$ and $\Dij{31}$ stand, respectively, for the radial transport and bootstrap current monoenergetic coefficients. Their precise definition and normalization is given in \cite{escoto2024monkes} but for this work is sufficient to know that, for fixed collisionality $\hat{\nu}$ and radial electric field ${E}_r$, the monoenergetic coefficients $\Dij{ij}$ encapsulate the dependence on the magnetic configuration of neoclassical transport in a given flux-surface. As until very recently \cite{escoto2024monkes} the inclusion of the $\Dij{31}$ coefficient in the optimization loop was practically impossible, the bootstrap current has traditionally been optimized indirectly. That is, some proxies which vanish for exactly QI configurations are used and then one \textit{hopes} that minimizing them will also minimize $|\Dij{31}|$. However, with this approach, one cannot guarantee that reducing the proxies will translate in a sufficient minimization of $|\Dij{31}|$. Moreover, the indirect approach does not allow to optimize taking into account the effect of the bootstrap current on the magnetic configuration and its neoclassical properties. For stellarators which are sufficiently close to quasi-symmetry \cite{Landreman_SelfConsistent}, optimization can be done in a self-consistent manner using analytical formulae for the bootstrap current \cite{Redl_Bootstrap} that are accurate and fast to compute. Until the development of the fast neoclassical code {\MONKES} this was not practically feasible for general magnetic geometry. For non-symmetric configurations, analytical formulae for low collisionality are available (e.g. \cite{Shaing_Callen}), but these expressions are known to be inaccurate \cite{Landreman_SelfConsistent} and the convergence of finite collisionality results to the asymptotical collisionless limit is not guaranteed \cite{albert2024convergencebootstrapcurrentshaingcallen}.

 
 In \cite{Sanchez_2023}, a selection of new and standard proxies for quasi-isodynamicity is proposed, which allowed to obtain the \qmarks{flat-mirror} \cite{velasco2023robust} nearly QI configuration CIEMAT-QI4. In order to evaluate how efficient the optimization strategy was for minimizing neoclassical transport (and in particular the bootstrap current), we will use {\MONKES} to evaluate $\Dij{11}$ and $\Dij{31}$ for the database of magnetic configurations produced during the CIEMAT-QI4 optimization campaign. The efficiency of each proxy for indirect QI optimization will be assessed by investigating the correlation (or lack of it) between the proxy and $|\Dij{31}|$. It is important to remark that in the robust \qmarks{flat-mirror} strategy, many reactor relevant properties are optimized simultaneously. The key idea is not to focus on being extremely close to QI and instead tailor the magnetic field so that particles drift tangentially to the flux-surface. This trade-off facilitates to meet other reactor-relevant requirements that are not related to neoclassical transport e.g. magnetohydrodynamic stability. Therefore, this evaluation will clarify to what extent reducing these proxies translates into a reduction of $\Dij{31}$ when optimizing stellarators which are meant to be fusion reactor candidates. For the sake of clarity, we briefly recall those proxies which vanish for exactly omnigenous and QI fields.
 
 For radial transport, the so-called effective ripple $\epseff$ \cite{Nemov1999EvaluationO1} encapsulates neoclassical losses of bulk ions in the $1/\nu$ regime. For fast ions, the proxies $\Gamma_{\text{c}}$ \cite{NemovGammaC} 
 \begin{align}
 	\Gamma_{\text{c}}(s)
 	=
 	\frac{1}{\pi\sqrt{2}}
 	\mean*{
 	\int_{\Bmax^{-1}}^{\Bmin^{-1}}
 	\left(
 	\frac{
 		\overline{\vb*{v}_{\text{m}}\cdot\nabla s}
 	}
 	{
     	\overline{\vb*{v}_{\text{m}}\cdot\nabla\alpha}
    }
 	\right)^2 
 	\frac{B \dd{\lambda}}{\sqrt{1-\lambda B}} 
 	}
 	\label{eq:Gamma_c}
 \end{align}
 and its refinement $\Gamma_{\alpha}$ \cite{Velasco_2021} 
\begin{align}
	\Gamma_{\alpha}(s)
	=
	\frac{1}{\pi\sqrt{2}} 
	\left\langle 
		\int_{\Bmax^{-1}}^{\Bmin^{-1}}
		H \left(
		(\alpha_{\text{out}}-\alpha)\overline{\vb*{v}_{\text{m}}\cdot\nabla\alpha}
		\right) 
		\right.
		\nonumber
		\\
		\left.
		\times H \left(
		(\alpha-\alpha_{\text{in}})\overline{\vb*{v}_{\text{m}}\cdot\nabla\alpha}
		\right)
		\frac{B \dd{\lambda}}{\sqrt{1-\lambda B}} 
	\right\rangle
	\label{eq:Gamma_alpha}
\end{align}
are used. Here, $s$ is the flux-surface label, $\vb*{v}_{\text{m}}$ is the magnetic drift, $\lambda :=B^{-1}v_\perp^2/v^2 $ is the so called \qmarks{pitch-angle coordinate}, $\alpha:=\theta - \iota \zeta$ is a poloidal angle which labels field lines and its values $\alpha_{\text{in}}$ and $\alpha_{\text{out}}$ are defined in \cite{Velasco_2021} (their specific definitions are not relevant for this work). Here, $\theta$ and $\zeta$ denote respectively the poloidal and toroidal Boozer angles.

Several targets based on the shape of the isolines of $B$ in omnigenous configurations \cite{CaryPhysRevLett,Cary1997OmnigenityAQ} are also considered. In an omnigenous configuration (or for each well of those defined in \cite{Parra_2015}), all relative maxima and minima of $B$ have equal value. This implies that the variance of the relative maxima of $B$
\begin{align}
	\sigma^2(\Bmax^{\text{r}}) 
	:=
	\frac{1}{  N_\theta}
	\sum_{i=0}^{N_\theta-1}
	\left(
	\dfrac{
		B_{\text{M}}(\theta_i)
		- 
		B_{\text{M}}^{\text{mean}} 
	}{B_{00}}
	\right)^2
	,
\end{align}
and the variance of the relative minima
\begin{align}
	\sigma^2(\Bmin^{\text{r}}) 
	:=
	\frac{1}{  N_\theta}
	\sum_{i=0}^{N_\theta-1}
	\left(
	\dfrac{ 
		B_{\text{m}}(\theta_i)
		- 
		B_{\text{m}}^{\text{mean}}
	}{B_{00}}
	\right)^2
\end{align} 
vanish in a perfectly omnigenous configuration. Here, $B_{\text{M}}(\theta_i)=\max B(\theta_i,\zeta)$ and $B_{\text{m}}(\theta_i)=\min B(\theta_i,\zeta)$ for $0\le\zeta<2\pi/N_p$ are, respectively, the maximum and minimum values of $B$ in a poloidal equispaced grid $\theta_i =2\pi i/N_\theta$. The quantities $B_{\text{M}}^{\text{mean}} = \sum_{i=0}^{N_\theta-1} B_{\text{M}}(\theta_i)/N_\theta$ and $B_{\text{m}}^{\text{mean}} = \sum_{i=0}^{N_\theta-1} B_{\text{m}}(\theta_i)/N_\theta$ are, respectively, the mean values of $\{B_{\text{M}}(\theta_i)\}_{i=0}^{N_\theta-1}$ and $\{B_{\text{m}}(\theta_i)\}_{i=0}^{N_\theta-1}$.

In a QI stellarator, stellarator symmetry \cite{DEWAR1998275} implies that the maximum value of $B$ in the flux-surface must be attained at the beginning or the centre of the field period along a curve that closes poloidally. Thus, stellarator symmetry implies that the isoline $B=\Bmax$ must match either the curve $\zeta=0$ or $\zeta=\pi/\Nfp$, where $\Nfp$ is the number of field periods of the device. However, redefining the beginning of the field period (i.e. mapping $\zeta\mapsto \zeta -\pi/\Nfp$) permits to agglutinate both cases in the case $\zeta=0$. Thus, specifically for obtaining (stellarator-symmetric) QI configurations, the variance of $B$ at $\zeta=0$ is considered
\begin{align}
\sigma^2(B(\theta,0)) 
:=
\frac{1}{N_\theta}
\sum_{i=0}^{N_\theta-1}
\left(
\dfrac{
	B(\theta_i,0)
	-
	B_0^{\text{mean}}
}{ B_{00} }
\right)^2 
,
\end{align}  
where $B_0^{\text{mean}} = \sum_{i=0}^{N_\theta-1} B(\theta_i,0)/N_\theta$. Note that for a perfectly QI stellarator-symmetric magnetic field $\sigma^2(B(\theta,0)) $ vanishes but, by itself, the nullity of $\sigma^2(B(\theta,0)) $ does not guarantee that the curve $\zeta=0$ coincides with the isoline $B=\Bmax$.

For the neoclassical transport evaluation, we select a value of low collisionality $\hat{\nu}:=\nu(v)/v=\num{e-5}$ $\text{m}^{-1}$ and two of the radial electric field $\widehat{E}_r:=E_r/v \in \{0,\num{e-3}\}$ $\text{kV}\cdot \text{s}/\text{m}^2$. Here, $\nu$ is the collision frequency and $v$ is the speed. The cases with zero radial electric field are in the $1/\nu$ regime and those with finite $\widehat{E}_r$ are in the $\sqrt{\nu}$-$\nu$ regime \cite{dherbemont2022}. For each pair $(\hat{\nu},\widehat{E}_r)$, we calculate the monoenergetic transport coefficients $\Dij{11}$ and $\Dij{31}$ of each configuration from the large database ($\sim 10^3$) of configurations using {\MONKES}. In order to compare different magnetic configurations, we normalize the monoenergetic coefficients as in \cite{Beidler_2011} (further details in \ref{sec:Appendix_monoenergetic_normalization}) and we denote them by $D_{ij}^*$.
 
 \begin{figure*}
 	\centering
 	\tikzsetnextfilename{D31_vs_D11_eps_0100_Er_0}
 	\begin{subfigure}[t]{0.45\textwidth}	
 		\includegraphics{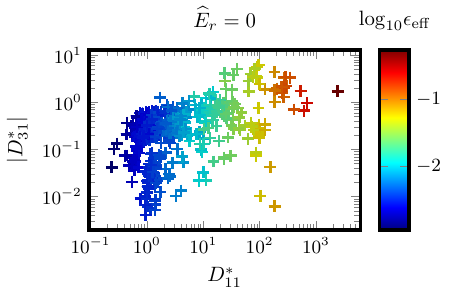}
 		\caption{}
 		\label{subfig:D31_vs_D11_eps_0100_Er_0}
 	\end{subfigure} 
 	\tikzsetnextfilename{D31_vs_D11_eps_0100_Er_1e-3}
 	\begin{subfigure}[t]{0.45\textwidth}	
        \includegraphics{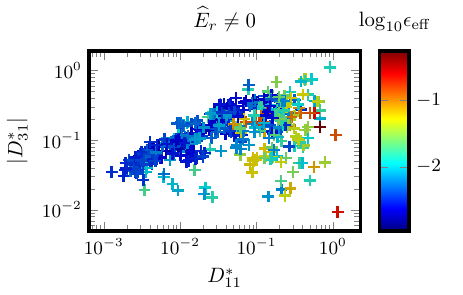}
 		\caption{}
 		\label{subfig:D31_vs_D11_eps_0100_Er_1e-3}
 	\end{subfigure} 
 	
 	\tikzsetnextfilename{D31_vs_D11_eps_0100_Er_0_restricted}
 	\begin{subfigure}[t]{0.45\textwidth}	
 		\includegraphics{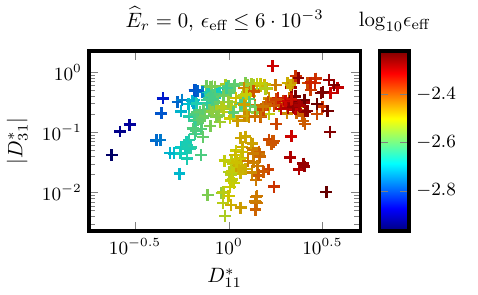}
 		\caption{}
 		\label{subfig:D31_vs_D11_eps_0100_Er_0_restricted}
 	\end{subfigure} 
 	\tikzsetnextfilename{D31_vs_D11_eps_0100_Er_1e-3_restricted}
 	\begin{subfigure}[t]{0.45\textwidth}	
 		\includegraphics{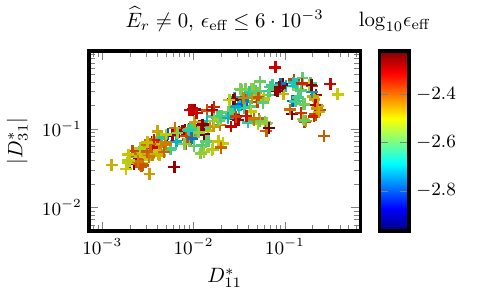}
 		\caption{}
 		\label{subfig:D31_vs_D11_eps_0100_Er_1e-3_restricted}
 	\end{subfigure}

 	\tikzsetnextfilename{D31_vs_eps_0100_Er_0}
 	\begin{subfigure}[t]{0.45\textwidth}	
 		\includegraphics{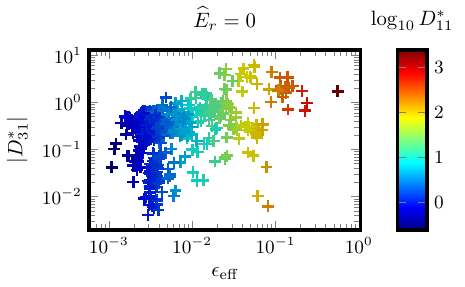}
 		\caption{}
 		\label{subfig:D31_vs_eps_0100_Er_0}
 	\end{subfigure} 
 	\tikzsetnextfilename{D31_vs_eps_0100_Er_1e-3}
 	\begin{subfigure}[t]{0.45\textwidth}	
 		\includegraphics{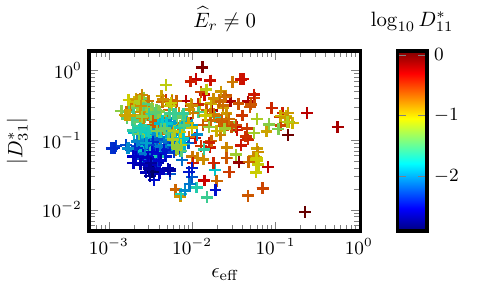}
 		\caption{}
 		\label{subfig:D31_vs_eps_0100_Er_1e-3}
 	\end{subfigure}      
 	\caption{Relation of the radial transport $D_{11}^*$ and bootstrap current $D_{31}^*$ coefficients with $\epseff$.}
 	\label{fig:Correlation_epseff}
 \end{figure*} 

 \begin{figure*}
 	\centering
 	\tikzsetnextfilename{D31_vs_D11_KN_VBB_0100_Er_0}
 	\begin{subfigure}[t]{0.45\textwidth}	 		
 		\includegraphics{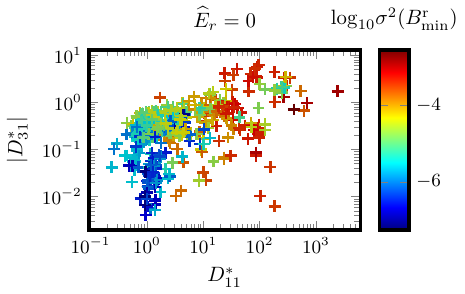}
 		\caption{}
 		\label{subfig:D31_vs_D11_KN_VBB_0100_Er_0}
 	\end{subfigure}     
 	\tikzsetnextfilename{D31_vs_D11_KN_VBB_0100_Er_1e-3}	
 	\begin{subfigure}[t]{0.45\textwidth}
 		\includegraphics{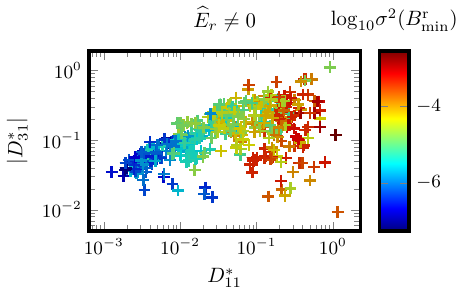}
 		\caption{}
 		\label{subfig:D31_vs_D11_KN_VBB_0100_Er_1e-3}
 	\end{subfigure} 

 	\tikzsetnextfilename{D31_vs_KN_VBB_0100_Er_0}	
 	\begin{subfigure}[t]{0.45\textwidth}	 		
 		\includegraphics{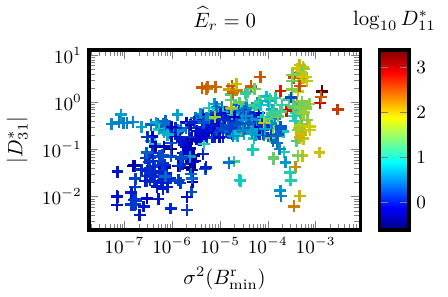}
 		\vspace{0.022cm} 
 		\caption{}
 		\label{subfig:D31_vs_KN_VBB_0100_Er_0}
 	\end{subfigure}     
 	\tikzsetnextfilename{D31_vs_KN_VBB_0100_Er_1e-3}	
 	\begin{subfigure}[t]{0.45\textwidth}	 		
 		\includegraphics{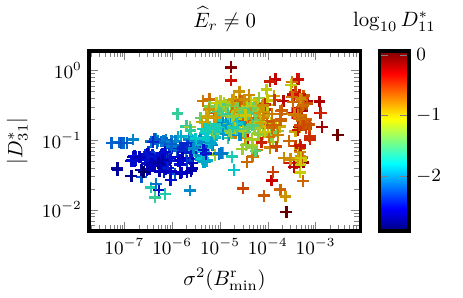}
 		\caption{}
 		\label{subfig:D31_vs_KN_VBB_0100_Er_1e-3}
 	\end{subfigure} 

 	\tikzsetnextfilename{D31_vs_KN_VBB_0100_Er_0_restricted}	
 	\begin{subfigure}[t]{0.45\textwidth}	 		
 		\includegraphics{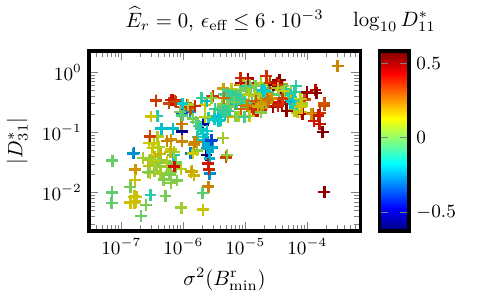}
 		\caption{}
 		\label{subfig:D31_vs_KN_VBB_0100_Er_0_restricted}
 	\end{subfigure}     
 	\tikzsetnextfilename{D31_vs_KN_VBB_0100_Er_1e-3_restricted}	
 	\begin{subfigure}[t]{0.45\textwidth}	 		
 		\includegraphics{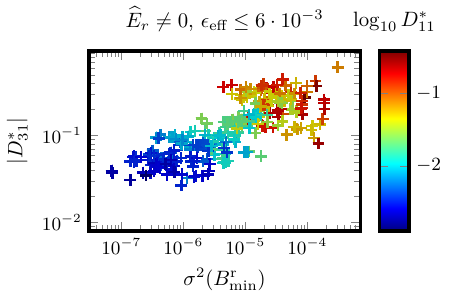}
 		\caption{}
 		\label{subfig:D31_vs_KN_VBB_0100_Er_1e-3_restricted}
 	\end{subfigure} 
 	\caption{Relation of the radial transport $D_{11}^*$ and bootstrap current $D_{31}^*$ coefficients with $\sigma^2(\Bmin^{\text{r}})$.} 
 	\label{fig:Correlation_VBB}
 \end{figure*}

 \begin{figure*}
 	
 	\tikzsetnextfilename{D31_vs_D11_KN_VB0_0100_Er_0}	
 	\begin{subfigure}[t]{0.45\textwidth}	
 		\includegraphics{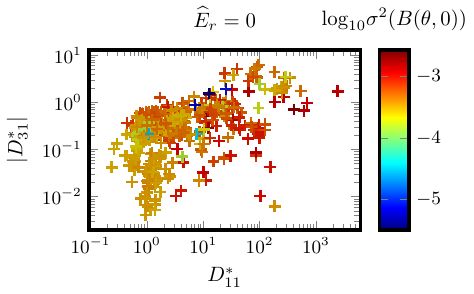}
 		\caption{}
 		\label{subfig:D31_vs_D11_KN_VB0_0100_Er_0}
 	\end{subfigure}
 	\tikzsetnextfilename{D31_vs_D11_KN_VB0_0100_Er_1e-3}	
 	\begin{subfigure}[t]{0.45\textwidth}	
 		\includegraphics{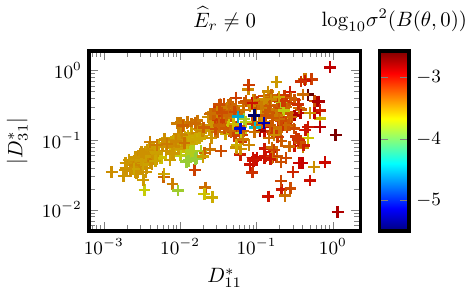}
 		\caption{}
 		\label{subfig:D31_vs_D11_KN_VB0_0100_Er_1e-3}
 	\end{subfigure} 
 	
 	\tikzsetnextfilename{D31_vs_KN_VB0_0100_Er_0}	
 	\begin{subfigure}[t]{0.45\textwidth}	
 		\includegraphics{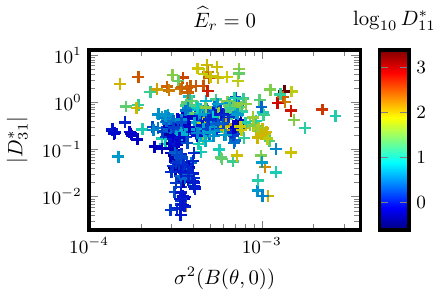}
 		\vspace{0.022cm} 
 		\caption{}
 		\label{subfig:D31_vs_KN_VB0_0100_Er_0}
 	\end{subfigure}
 	\tikzsetnextfilename{D31_vs_KN_VB0_0100_Er_1e-3}	
 	\begin{subfigure}[t]{0.45\textwidth}	
 		\includegraphics{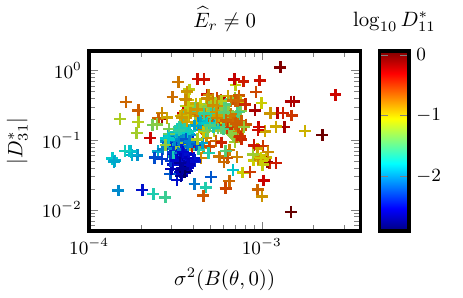}
 		\caption{}
 		\label{subfig:D31_vs_KN_VB0_0100_Er_1e-3}
 	\end{subfigure}

 	\tikzsetnextfilename{D31_vs_KN_VB0_0100_Er_0_restricted}	
 	\begin{subfigure}[t]{0.45\textwidth}	
 		\includegraphics{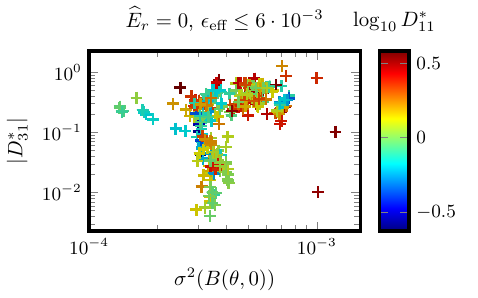}
 		\caption{}
 		\label{subfig:D31_vs_KN_VB0_0100_Er_0_restricted}
 	\end{subfigure}
 	\tikzsetnextfilename{D31_vs_KN_VB0_0100_Er_1e-3_restricted}	
 	\begin{subfigure}[t]{0.45\textwidth}	
 		\includegraphics{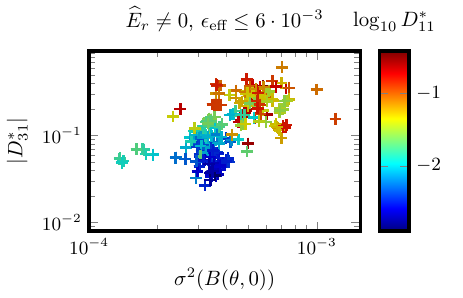}
 		\caption{}
 		\label{subfig:D31_vs_KN_VB0_0100_Er_1e-3_restricted}
 	\end{subfigure} 
 	\caption{Relation of the radial transport $D_{11}^*$ and bootstrap current $D_{31}^*$ coefficients with $\sigma^2(B(\theta,0))$.}
 	\label{fig:Correlation_VB0}
 \end{figure*} 
 
 \begin{figure*}     
 	\tikzsetnextfilename{D31_vs_D11_KN_GMC_0100_Er_0}	
 	\begin{subfigure}[t]{0.45\textwidth}	 
 		\includegraphics{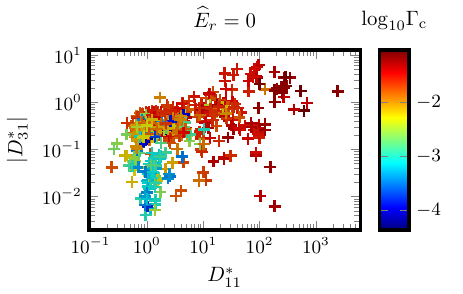}
 		\caption{}
 		\label{subfig:D31_vs_D11_KN_GMC_0100_Er_0}
 	\end{subfigure}
 	\tikzsetnextfilename{D31_vs_D11_KN_GMA_0100_Er_0}	
 	\begin{subfigure}[t]{0.45\textwidth}	 
        \includegraphics{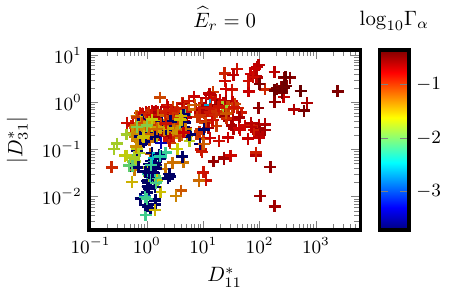}
 		\caption{}
 		\label{subfig:D31_vs_D11_KN_GMA_0100_Er_0}
 	\end{subfigure}

 	\tikzsetnextfilename{D31_vs_KN_GMC_0100_Er_0}	
 	\begin{subfigure}[t]{0.45\textwidth}	 
 		\includegraphics{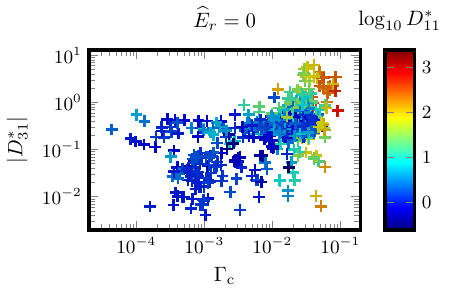}
 		\caption{}
 		\label{subfig:D31_vs_KN_GMC_0100_Er_0}
 	\end{subfigure}
 	\tikzsetnextfilename{D31_vs_KN_GMA_0100_Er_0}	
 	\begin{subfigure}[t]{0.45\textwidth}	 
 		\includegraphics{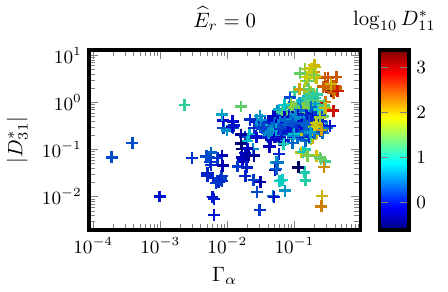}
 		\vspace{0.022cm} 
 		\caption{}
 		\label{subfig:D31_vs_KN_GMA_0100_Er_0}
 	\end{subfigure}

 	\tikzsetnextfilename{D31_vs_KN_GMC_0100_Er_0_restricted}	
 	\begin{subfigure}[t]{0.45\textwidth}	 
 		\includegraphics{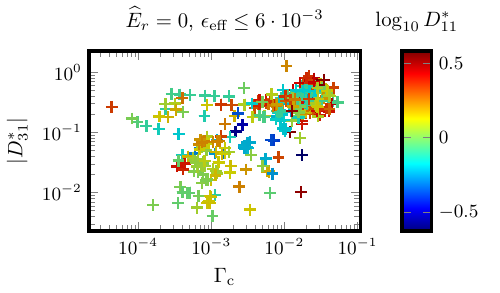}
 		\caption{}
 		\label{subfig:D31_vs_KN_GMC_0100_Er_0_restricted}
 	\end{subfigure}
 	\tikzsetnextfilename{D31_vs_KN_GMA_0100_Er_0_restricted}	
 	\begin{subfigure}[t]{0.45\textwidth}	 
 		\includegraphics{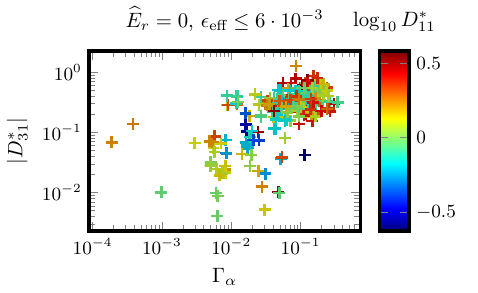}
 		\caption{}
 		\label{subfig:D31_vs_KN_GMA_0100_Er_0_restricted}
 	\end{subfigure}

 	\caption{Relation of the radial transport $D_{11}^*$ and bootstrap current $D_{31}^*$ coefficients with $\GammaC$ and $\GammaAlpha$ for $\widehat{E}_r=0$.} 
 	\label{fig:Correlation_GMC_GMA_Er_0}
 \end{figure*}

 \begin{figure*}
 	\centering
 	\tikzsetnextfilename{D31_vs_D11_KN_GMC_0100_Er_1e-3}	
 	\begin{subfigure}[t]{0.45\textwidth}	 
 		\includegraphics{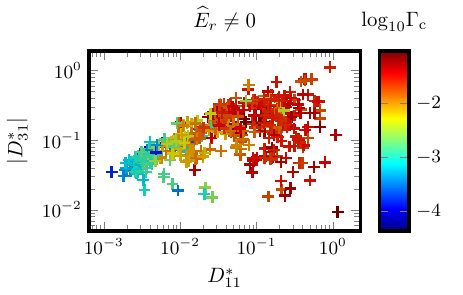}
 		\caption{}
 		\label{subfig:D31_vs_D11_KN_GMC_0100_Er_1e-3}
 	\end{subfigure}
 	\tikzsetnextfilename{D31_vs_D11_KN_GMA_0100_Er_1e-3}	
 	\begin{subfigure}[t]{0.45\textwidth}	 
 		\includegraphics{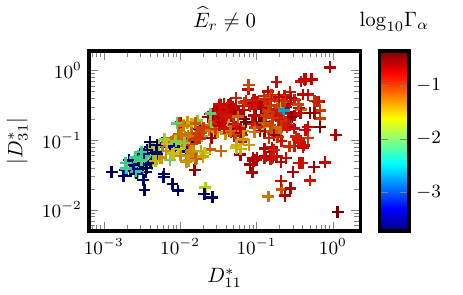}
 		\caption{}
 		\label{subfig:D31_vs_D11_KN_GMA_0100_Er_1e-3}
 	\end{subfigure}

 	\tikzsetnextfilename{D31_vs_KN_GMC_0100_Er_1e-3}	
 	\begin{subfigure}[t]{0.45\textwidth}	 
 		\includegraphics{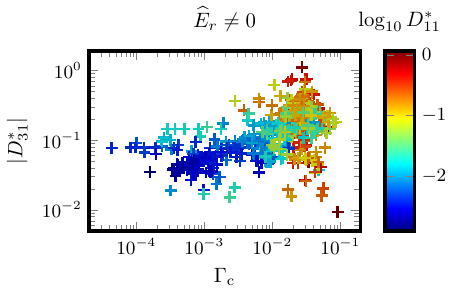}
 		\caption{}
 		\label{subfig:D31_vs_KN_GMC_0100_Er_1e-3}
 	\end{subfigure}
 	\tikzsetnextfilename{D31_vs_KN_GMA_0100_Er_1e-3}	
 	\begin{subfigure}[t]{0.45\textwidth}	 
 		\includegraphics{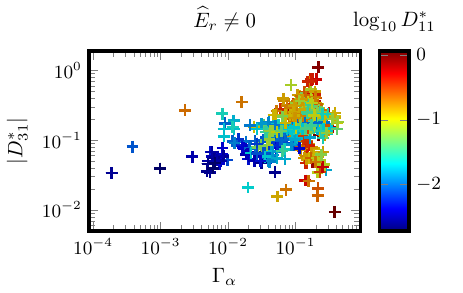}
 		\caption{}
 		\label{subfig:D31_vs_KN_GMA_0100_Er_1e-3}
 	\end{subfigure}

 	\tikzsetnextfilename{D31_vs_KN_GMC_0100_Er_1e-3_restricted}	
 	\begin{subfigure}[t]{0.45\textwidth}	 
 		\includegraphics{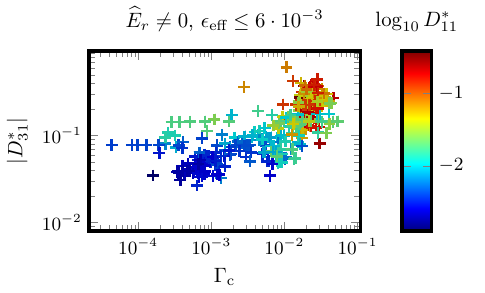}
 		\caption{}
 		\label{subfig:D31_vs_KN_GMC_0100_Er_1e-3_restricted}
 	\end{subfigure}
 	\tikzsetnextfilename{D31_vs_KN_GMA_0100_Er_1e-3_restricted}	
 	\begin{subfigure}[t]{0.45\textwidth}	 
 		\includegraphics{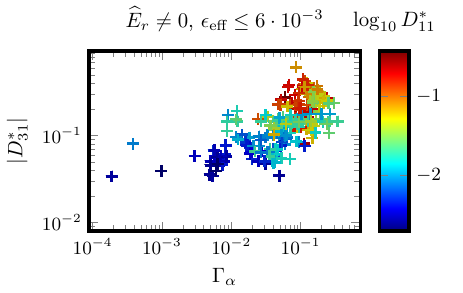}
 		\caption{}
 		\label{subfig:D31_vs_KN_GMA_0100_Er_1e-3_restricted}
 	\end{subfigure}

 	\caption{Relation of the radial transport $D_{11}^*$ and bootstrap current $D_{31}^*$ coefficients with $\GammaC$ and $\GammaAlpha$ for $\widehat{E}_r\ne 0$.} 
 	\label{fig:Correlation_GMC_GMA_Er_1e-3}
 \end{figure*}
 
In figures \ref{subfig:D31_vs_D11_eps_0100_Er_0} and \ref{subfig:D31_vs_D11_eps_0100_Er_1e-3} the result of the neoclassical evaluation for the database of the CIEMAT-QI4 campaign is shown, along with the value of $\epseff$ in colours. Each point on the plane $D_{11}^*-|D_{31}^*|$ corresponds to a different configuration with a particular value of $\epseff$. Thus, configurations closer to being QI are located near the bottom left corner of these plots. Figure \ref{subfig:D31_vs_D11_eps_0100_Er_0} shows that, in the absence of radial electric field, configurations which were optimized for having small $D_{11}^*$ (equivalently $\epseff$) not necessarily had small bootstrap current coefficient. On the other hand, in the presence of a finite $\widehat{E}_r$, we can see from figure  \ref{subfig:D31_vs_D11_eps_0100_Er_1e-3} that minimizing radial transport entailed a minimization of $D_{31}^*$. In the complete database shown in figures \ref{subfig:D31_vs_D11_eps_0100_Er_0} and \ref{subfig:D31_vs_D11_eps_0100_Er_1e-3} there are many configurations corresponding to the initial stages of the optimization campaign and therefore, are not sufficiently optimized. As $\epseff$ is typically used as an indicator of overall radial neoclassical transport optimization, for inspecting potential correlations, it is useful to filter out non optimized configurations. When we restrict the database to those configurations with $ \epseff \lesssim \num{6e-3} $ for the case of $\widehat{E}_r=0$, the results shown in figure \ref{subfig:D31_vs_D11_eps_0100_Er_0_restricted} suggest a trade-off between $D_{11}^*$ and $|D_{31}^*|$. Conversely, note from figure  \ref{subfig:D31_vs_D11_eps_0100_Er_1e-3_restricted} that those configurations optimized to have $\epseff \lesssim \num{6e-3}$ cluster around a straight line of the $D_{11}^*-|D_{31}^*|$ plane. This clustering indicates a moderate correlation between $D_{11}^*$ and $|D_{31}^*|$ for sufficiently optimized configurations in the presence of a non zero radial electric field. The distribution of colours in figures \ref{subfig:D31_vs_D11_eps_0100_Er_0} and \ref{subfig:D31_vs_D11_eps_0100_Er_1e-3} also reveals that there is no correlation between $\epseff$ and $|D_{31}^*|$. This lack of correlation can be seen in more detail in figures \ref{subfig:D31_vs_eps_0100_Er_0} and \ref{subfig:D31_vs_eps_0100_Er_1e-3}, where the projection of the data onto the $|D_{31}^*|-\epseff$ plane is shown and the value of $D_{11}^*$ is represented in colours. Note that, for both values of $\widehat{E}_r$, those configurations that display simultaneously small levels of parallel and radial neoclassical transport are those with minimum $\epsilon_{\text{eff}}$. However, reducing $\epsilon_{\text{eff}}$ does not guarantee a reduction of the $D_{31}^*$ coefficient. For $\epseff\sim \num{2e-3}$ and $\widehat{E}_r=0$, we can see in figure \ref{subfig:D31_vs_eps_0100_Er_0} that the bootstrap current coefficient can range in an interval of almost three orders of magnitude, from $|D_{31}^*| \sim \num{e-3}$ to $|D_{31}^*| \sim 1$. For the case with finite $\widehat{E}_r$ shown in figure \ref{subfig:D31_vs_eps_0100_Er_1e-3}, the situation is similar but with a narrower interval of $|D_{31}^*|$. For $\epseff\sim \num{2e-3}$, the bootstrap current coefficient can change an order of magnitude, ranging between $|D_{31}^*| \sim \num{e-2}$ to $|D_{31}^*| \sim \num{e-1}$. This lack of correlation is unsurprising as reducing the effective ripple guarantees proximity to omnigenity, which is a necessary but not sufficient condition for quasi-isodynamicity. Finally, the fact that in the $1/\nu$ regime $D_{11}^* \propto \epseff^{3/2}/\hat{\nu}$ can be seen from figures \ref{subfig:D31_vs_D11_eps_0100_Er_0} and \ref{subfig:D31_vs_eps_0100_Er_0}. Note that the distribution of points and colour in figures \ref{subfig:D31_vs_D11_eps_0100_Er_0} and \ref{subfig:D31_vs_eps_0100_Er_0} is almost identical. Of course, this nearly perfect correlation is not preserved for the $\sqrt{\nu}$-$\nu$ regime, as shown in figures \ref{subfig:D31_vs_D11_eps_0100_Er_1e-3} and \ref{subfig:D31_vs_eps_0100_Er_1e-3}. This was expected as particle trajectories that cause the $\sqrt{\nu}$-$\nu$ flux are quite different from those that generate the $1/\nu$ flux.


In figure \ref{fig:Correlation_VBB} the relation between $\sigma^2(\Bmin^{\text{r}})$ and the monoenergetic coefficients during the optimization campaign is shown. From figure \ref{subfig:D31_vs_D11_KN_VBB_0100_Er_0}, we can see that the smallest values of $\sigma^2(\Bmin^{\text{r}})$ cluster around the smallest values of $D_{11}^*$ and in the range of bootstrap current coefficient $\num{e-2}\lesssim D_{31}^*\lesssim \num{e-1}$. This suggests a slight correlation between $D_{31}^*$ and the variance $\sigma^2(\Bmin^{\text{r}})$. However, when inspecting this correlation in more detail in figure \ref{subfig:D31_vs_KN_VBB_0100_Er_0}, we can see that for very small values of $\sigma^2(\Bmin^{\text{r}})\lesssim 10^{-6}$, $D_{31}^*$ can vary almost two orders of magnitude, even if $D_{11}^*$ is also small. This simply indicates that it is possible to have a large deviation from quasi-isodynamicity even if $\sigma^2(\Bmin^{\text{r}})$ is close to zero. In figure \ref{subfig:D31_vs_KN_VBB_0100_Er_0_restricted} we have filtered out those configurations with $\epseff>\num{6e-3}$ and the slight correlation for sufficiently optimized configurations (in terms of the $\epseff$) is apparent, but far from ideal as $|D_{31}^*|$ can vary two orders of magnitude for $\sigma^2(\Bmin^{\text{r}})\sim\num{5e-7}$. The variability in the value of $|D_{31}^*|$ was expected, as this proxy is meant for approaching general omnigenity. As expected, there seems to be a trade-off between radial and parallel transport as the configurations with the smallest value of $|D_{31}^*|$ do not have the smallest values of $D_{11}^*$ aswell. For the case with radial electric field the correlation seems to be stronger. We can see in figure \ref{subfig:D31_vs_D11_KN_VBB_0100_Er_1e-3} that the smallest values of $\sigma^2(\Bmin^{\text{r}})$ are clustered very close to the left inferior corner in the $D_{11}^*-|D_{31}^*|$ plane. Indeed, for the smallest values of $\sigma^2(\Bmin^{\text{r}})$, we can see in figure \ref{subfig:D31_vs_KN_VBB_0100_Er_1e-3} that $|D_{31}^*|\lesssim \num{e-1}$. As shown in figure \ref{subfig:D31_vs_KN_VBB_0100_Er_1e-3_restricted}, the correlation is more evident for configurations optimized to have $\epseff\le\num{6e-3}$. The results suggest that for the case with finite radial electric field there is a moderate correlation between $\sigma^2(\Bmin^{\text{r}})$ and $|D_{31}^*|$. However, from the horizontal spread of the points shown in figure \ref{subfig:D31_vs_KN_VBB_0100_Er_1e-3_restricted}, we can conclude that minimizing $\sigma^2(\Bmin^{\text{r}})$ can be very inefficient for reducing $|D_{31}^*|$.

In figure \ref{fig:Correlation_VB0} the relation between the monoenergetic coefficients and $\sigma^2(B(\theta,0))$ is shown. It is immediate to see from \ref{subfig:D31_vs_D11_KN_VB0_0100_Er_0} that, for $\widehat{E}_r=0$, there is no correlation between $\sigma^2(B(\theta,0))$ and the bootstrap current coefficient. Inspecting the lack of correlation in more detail in \ref{subfig:D31_vs_KN_VB0_0100_Er_0} we confirm that minimizing the variance from $\sigma^2(B(\theta,0))\sim \num{3e-4}$ to $\sigma^2(B(\theta,0))\sim\num{1e-4}$ can increase substantially the bootstrap current coefficient, even if $D_{11}^*$ is kept below 1. If we filter out configurations with $\epseff>\num{6e-3}$, as shown in figure \ref{subfig:D31_vs_KN_VB0_0100_Er_0_restricted}, this behaviour is confirmed and the results suggest that the simultaneous minimization of $\sigma^2(B(\theta,0))$ and $\epseff$ ($D_{11}^*$) is done at the expense of increasing $|D_{31}^*|$. Note from figure \ref{subfig:D31_vs_KN_VB0_0100_Er_0} that configurations with smaller levels of radial and parallel transport cluster at intermediate values of the variance $\sigma^2(B(\theta,0))$. This behaviour persists even for configurations with sufficiently optimized effective ripple, as shown in \ref{subfig:D31_vs_KN_VB0_0100_Er_0_restricted}. For the case with finite radial electric field, from figure \ref{subfig:D31_vs_D11_KN_VB0_0100_Er_1e-3}, we can see no appreciable correlation between $\sigma^2(B(\theta,0))$ and $|D_{31}^*|$. In figure \ref{subfig:D31_vs_KN_VB0_0100_Er_1e-3} we can see that configurations with small values of $D_{11}^*$ and $|D_{31}^*|$ cluster near the left of the plot, but still without strong correlation. When we filter configurations which are not sufficiently optimized in terms of $\epseff$, the results shown in figure \ref{subfig:D31_vs_KN_VB0_0100_Er_1e-3_restricted} suggest a mild correlation between $|D_{31}^*|$ and $\sigma^2(B(\theta,0))$ for $\widehat{E}_r\ne 0$. However, it is very far from ideal as for $\sigma^2(B(\theta,0))\sim\num{3e-4}$ the radial transport and bootstrap current coefficient can vary, respectively, two and one orders of magnitude. The inadequacy of $\sigma^2(B(\theta,0))$ for minimizing $|D_{31}^*|$ (even for configurations with small $\epseff$) is surprising as this is the only proxy specific for optimizing QI configurations and naively one would expect a better correlation. Finally, we point out that for the database considered, the proxies $\sigma^2(B(\theta,0))$ and $\sigma^2(\Bmax^{\text{r}}) $ are roughly equivalent and therefore we omit the results for the latter. This equivalency between the two proxies can be seen from figure \ref{fig:Correlation_VBM} in \ref{sec:Appendix_VBM}, which is very similar to figure \ref{fig:Correlation_VB0}.


Finally, we compare the relation of the monoenergetic coefficients with the fast ion proxies $\GammaC$ and $\GammaAlpha$. In figure \ref{fig:Correlation_GMC_GMA_Er_0}, the case for zero radial electric field is shown. Note from figure \ref{subfig:D31_vs_D11_KN_GMC_0100_Er_0} that configurations with the smallest values of $\GammaC$ do not cluster near the left inferior corner but on values $|D_{31}^*|\sim\num{3e-1}$. On the other hand, as figure \ref{subfig:D31_vs_D11_KN_GMA_0100_Er_0} shows, configurations with the smaller levels of parallel and radial transport also have the smallest values of $\GammaAlpha$. This difference suggests a slightly better correlation between $|D_{31}^*|$ and $\GammaAlpha$ than between $|D_{31}^*|$ and $\GammaC$. Inspecting this difference further, we can see in figure \ref{subfig:D31_vs_KN_GMC_0100_Er_0} that for $\GammaC$ there is an horizontal branch along which we can reduce $\GammaC$ but not $|D_{31}^*|$ and its value is not small ($|D_{31}^*|>10^{-1}$). As shown in figure \ref{subfig:D31_vs_KN_GMA_0100_Er_0}, this is not the case for $\GammaAlpha$ which seems to have a mild correlation with $|D_{31}^*|$. This difference in the behaviour persists even for configurations with low values of the effective ripple. From figure \ref{subfig:D31_vs_KN_GMC_0100_Er_0_restricted} we can see that the horizontal branch of $\GammaC$ is still present for configurations with low value of $\epseff$. Conversely, in figure \ref{subfig:D31_vs_KN_GMA_0100_Er_0_restricted} we can see that the correlation between $|D_{31}^*|$ and $\GammaAlpha$ is more pronounced for configurations with low $\epseff$. The case with finite radial electric field is shown in figure \ref{fig:Correlation_GMC_GMA_Er_1e-3}. For the case $\widehat{E}_r \ne 0$, the discussion is similar to the case without radial electric field. These numerical results suggest that in order to obtain a finite but small bootstrap current, it is more important to have contours of the second adiabatic invariant $J$ which close poloidally and do not deviate much from flux-surfaces rather than exactly matching them. Specifically, for an approximately omnigenous configuration, reducing $\GammaC$ implies aligning all $J$ isosurfaces with flux-surfaces. On the other hand, minimizing $\GammaAlpha$ entails an alignment of those constant $J$ surfaces which deviate the most from flux-surfaces, but not all of them. A different (although non exclusive) possibility could be that, enforcing $J$ contours to be closed surfaces by minimizing $\GammaAlpha$ facilitates achieving the maximum$-J$ \cite{Rosenbluth_max_J,Helander_max_J} property to a sufficient degree of approximation. If so, the optimizer would be able to focus on minimizing other quantities, such as $\sigma^2(\Bmin^{\text{r}})$ to reduce $|D_{31}^*|$.

%

To summarize, in the light of these results, we can conclude that, although effective for obtaining nearly QI \qmarks{flat-mirror} configurations, the optimization strategy was not efficient for reducing the bootstrap current. The inefficiency of the indirect approach to minimize the $|D_{31}^*|$ coefficient is specially pronounced for the case without radial electric field. Thus, many intermediate configurations which are not sufficiently close from QI (in the sense of having too large $|D_{31}^*|$ or $D_{11}^*$) are produced during the optimization campaign. Apart from the imperfect correlation of the proxies used in indirect optimization, this inefficiency is probably enhanced by the multiple trade-offs that occur when many different requirements have to be met. It is reasonable to expect that a direct minimization of $|D_{31}^*|$, will be a much more efficient strategy for optimizing QI configurations.

\begingroup
\section{From quasi-isodynamicity to piecewise omnigenity} \label{sec:pwO_QI}
\captionsetup[sub]{skip=-1.75pt, margin=40pt}
\begin{figure}
	\centering
	\foreach \p in {20}
	{ 
		\foreach \w in {0.5, 0.9, 1.0, 1.3}
		{%
			\tikzsetnextfilename{pwQI_B_\w_\p}
			\begin{subfigure}[t]{0.23\textwidth} 
				\includegraphics{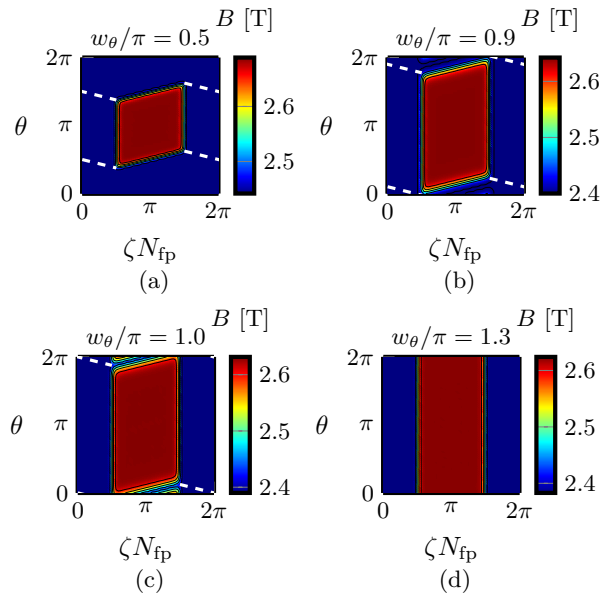}
				\caption{}
				\label{subfig:pwOMagneticField_pow_\p_wa\w_pi}
			\end{subfigure}%
		} 		
	} 
	\caption{Magnetic field strength $B$ of an approximately pwO magnetic field ($p=10$) for several values of $w_\theta$.}
	\label{fig:Magnetic_field_strength_pwO_QI}
\end{figure}

\begin{figure*}
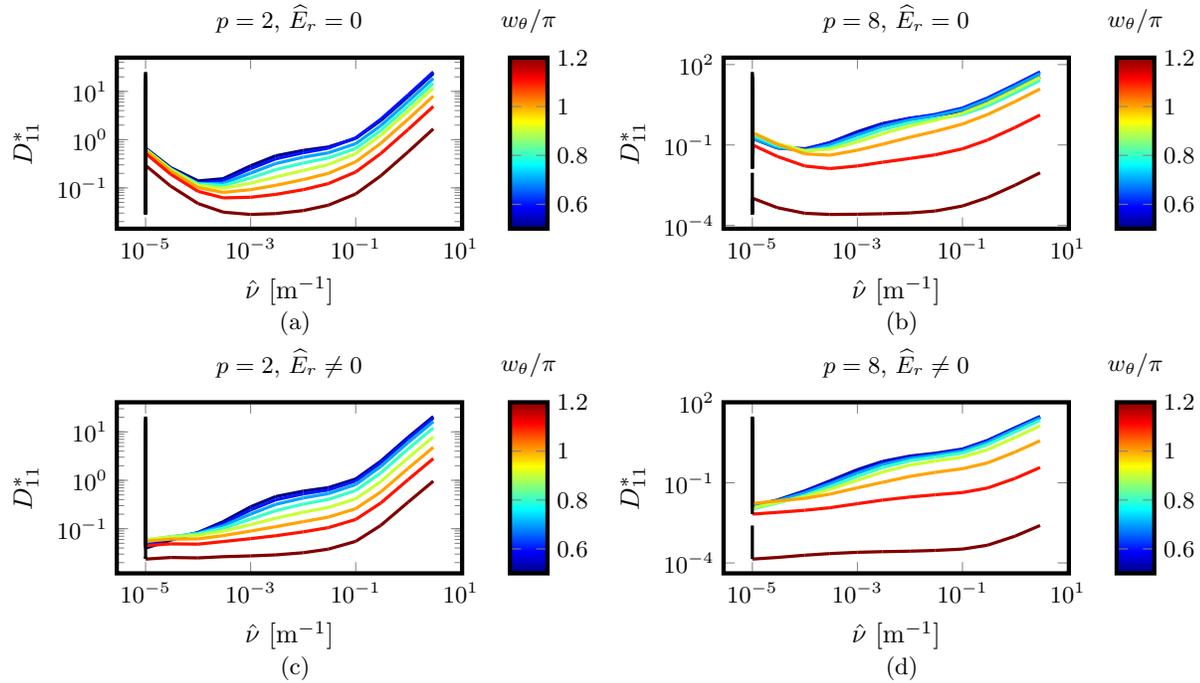
   
	\foreach \p in {4,16}
	{
		\tikzsetnextfilename{D11_vs_nu_pwQI_\p_Er_0}
		\begin{subfigure}[t]{0.45\textwidth}
			\includegraphics{D11_vs_nu_pwQI_\p_Er_0}
%
%
			\caption{}
			\label{subfig:D11_vs_nu_pwQI_\p_Er_0}
		\end{subfigure} 	     
	}	 
    
    \foreach \p in {4,16}
    { 
    	\tikzsetnextfilename{D11_vs_nu_pwQI_\p_Er_1e-3}
    	\begin{subfigure}[t]{0.45\textwidth}
    		\includegraphics{D11_vs_nu_pwQI_\p_Er_1e-3}
%
%
    		\caption{}
    		\label{subfig:D11_vs_nu_pwQI_\p_Er_1e-3}
    	\end{subfigure}     	
    }
    
	\caption{Radial transport $D_{11}^*$ and bootstrap current $D_{31}^*$ coefficients as functions of $\hat{\nu}$ and $w_\theta$ for $\widehat{E}_r = 0$}
	\label{fig:Dij_vs_nu_pwQI_Er_0}
\end{figure*}

\begin{figure*}
	\centering
	\begin{subfigure}[t]{0.45\textwidth}		
		\tikzsetnextfilename{D11_vs_pi_factor_nu_1e-5_Er_0}
		\includegraphics{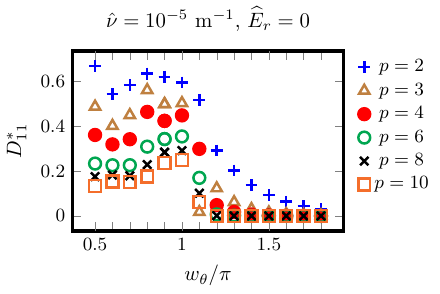}
		\caption{}
		\label{subfig:D11_vs_pi_factor_nu_1e-5}
	\end{subfigure}%
	\begin{subfigure}[t]{0.45\textwidth}		
		\tikzsetnextfilename{D11_vs_pi_factor_nu_1e-5_Er_1e-3}
		\includegraphics{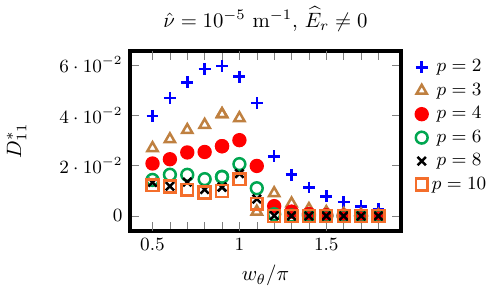}
		\caption{}
		\label{subfig:D11_vs_pi_factor_nu_1e-5_Er_1e-3}
	\end{subfigure}
	\caption{Radial transport coefficient $D_{11}^*$ as a function of $w_\theta$ and $p$ for $\hat{\nu}=\num{e-5}$ $\text{m}^{-1}$. (a) $\widehat{E}_r=0$ and (b) $\widehat{E}_r\ne0$. }
\end{figure*}
 
As a second application and demonstration of its capabilities, {\MONKES} will be used to study neoclassical transport in a novel family of optimized stellarator magnetic fields. Recently, the notion of \textit{piecewise omnigenity} has been introduced \cite{velasco2024piecewise}, broadening the configuration space of stellarators with minimal neoclassical energy transport. Piecewise omnigenous stellarators are omnigenous almost everywhere on the flux-surface, which means that $J$ is a flux-function piecewisely. Those zero measure regions where $J$ can vary on the flux-surface delimit different classes of trapped particles and therefore, transitions can occur when particles precess on the flux-surface due to drifts or when particles collide. Remarkably, as indicated in \cite{velasco2024piecewise}, these transitions do not contribute to radial transport in the $1/\nu$ regime. In regard to parallel transport, the bootstrap current produced by piecewise omnigenous stellarators is still an unexplored area. For future stellarator designs, it could be very helpful to find pwO magnetic fields which have not only reduced $\epseff$ but also a small bootstrap current. In this section, we will investigate neoclasically nearly pwO magnetic fields that are approximately QI. The objective is to identify which levels of approximate piecewise omnigenity and quasi-isodynamicity are necessary to have low levels of radial and parallel neoclassical transport. Incidentally, we will demonstrate that {\MONKES} can be used to study neoclassical transport in stellarator configurations which are extremely complicated in terms of the Fourier spectra of $B$. A simple pwO magnetic field can be modelled using an exponential \cite{velasco2024piecewise}

\begin{align}
	\widetilde{B}(\theta,\zeta)  
	& =
	\Bmin
	+
	(\Bmax-\Bmin)
	\exp(
	-
	\left(
	\frac{\zeta - \zeta_{\text{c}} }{w_\zeta}
	\right)^{2p}
	)
	\nonumber
	\\
	&
	\times
	\exp(
	-
	\left(
	\frac{\theta - \theta_{\text{c}} - t_\zeta(\zeta - \zeta_{\text{c}} )}{w_\theta}
	\right)^{2p}),
	\label{eq:Exponential_pwO}
\end{align}
along with the constraint to the rotational transform
\begin{align}
	\iota =  -t_\zeta \frac{\Nfp w_\zeta}{\pi - \Nfp w_\zeta} ,
	\label{eq:iota_pwO}
\end{align}
becoming exactly pwO in the limit $p\rightarrow \infty$. When $p\rightarrow \infty$, the isolines $\Bmin<\widetilde{B}<\Bmax$ are compressed in a single parallelogram of centre $(\theta_{\text{c}}, \zeta_{\text{c}})$ in the $(\theta,\zeta)$ plane. The four sides of this parallelogram are defined by the equations
\begin{align}
	 \theta - \theta_{\text{c}} & = \pm w_\theta + t_\zeta\left( \zeta - \zeta_{\text{c}}\right),
	 \label{eq:pwO_tilted_sides}
	 \\
	 \zeta - \zeta_{\text{c}} & = \pm w_\zeta.
	 \label{eq:pwO_vertical_sides}
\end{align}
Thus, the scalars $2w_\theta$ and $2w_\zeta<2\pi/\Nfp$ define, respectively, the poloidal and toroidal width of this parallelogram. The slope $t_\zeta$ defines the poloidal shear of the parallelogram, becoming a rectangle when $t_\zeta=0$. It is important to remark that the constraint (\ref{eq:iota_pwO}) guarantees that $J$ is a flux-function piecewisely for a field whose magnetic field strength is given by (\ref{eq:Exponential_pwO}) in the limit $p\rightarrow\infty$. We will precise this assertion later in this section.

In order to identify regions of the pwO parameter space with small $D_{11}^*$ and $|D_{31}^*|$ we will evaluate neoclassically approximately pwO magnetic fields obtained from a scan in $w_\theta$ for several values of finite $p$. The idea is to start from a value of $w_\theta$ for which the configuration is nearly pwO and increase it until it becomes nearly QI. In figure \ref{fig:Magnetic_field_strength_pwO_QI}, we illustrate the scan in $w_\theta$ for a fixed value of $p=10$ using the magnetic field strength $B$ of an approximately pwO field constructed in the manner instructed in \ref{sec:Appendix_pwO_B}. Note that in figure \ref{fig:Magnetic_field_strength_pwO_QI} we represent $B$ and not $\widetilde{B}$. We do this because the function $\widetilde{B}$ given by (\ref{eq:Exponential_pwO}) in the limit $p\rightarrow\infty$, by itself, can only define a stellarator-symmetric exactly pwO magnetic field strength $B$ for $w_\theta\le \pi-|t_\zeta| w_\zeta$. For a stellarator-symmetric exactly pwO field, at $w_\theta=\pi-|t_\zeta| w_\zeta$, two corners of the parallelogram are located at the poloidal positions $\theta=0$ and $\theta=2\pi$ and the remaining two somewhere in the interval $\theta \in (0,2\pi)$. An approximation to this situation is shown in figure \ref{subfig:pwOMagneticField_pow_20_wa0.9_pi}. If we increase $w_\theta $ beyond this point, the parallelogram does not fit in the domain $\theta \in [0,2\pi]$ and imposing the constraint (\ref{eq:iota_pwO}) no longer guarantees exact piecewise omnigenity. In order to increase $w_\theta$ and maintain approximate pwO, the parallelogram must \qmarks{grow} in the way shown in figure \ref{subfig:pwOMagneticField_pow_20_wa1.0_pi}. This behaviour cannot be obtained by simply increasing $w_\theta$ in the definition (\ref{eq:Exponential_pwO}). In addition, more complications arise when $p$ is finite. Nevertheless, as explained in \ref{sec:Appendix_pwO_B}, we can circumvent these complications and use the exponential function from equation (\ref{eq:Exponential_pwO}) and the constraint (\ref{eq:iota_pwO}), to construct a stellarator-symmetric approximately pwO field for different values of $w_\theta$ and finite $p$, including $w_\theta > \pi - |t_\zeta| w_\zeta$. The approximately pwO magnetic field has been constructed so that it resembles that of a flux-surface of Wendelstein 7-X KJM (further details in \ref{sec:Appendix_pwO_B}). The parameters required for defining the magnetic field are listed in tables \ref{tab:pwO_parameters} and \ref{tab:pwO_parameters_KJM}.

\begin{figure*}
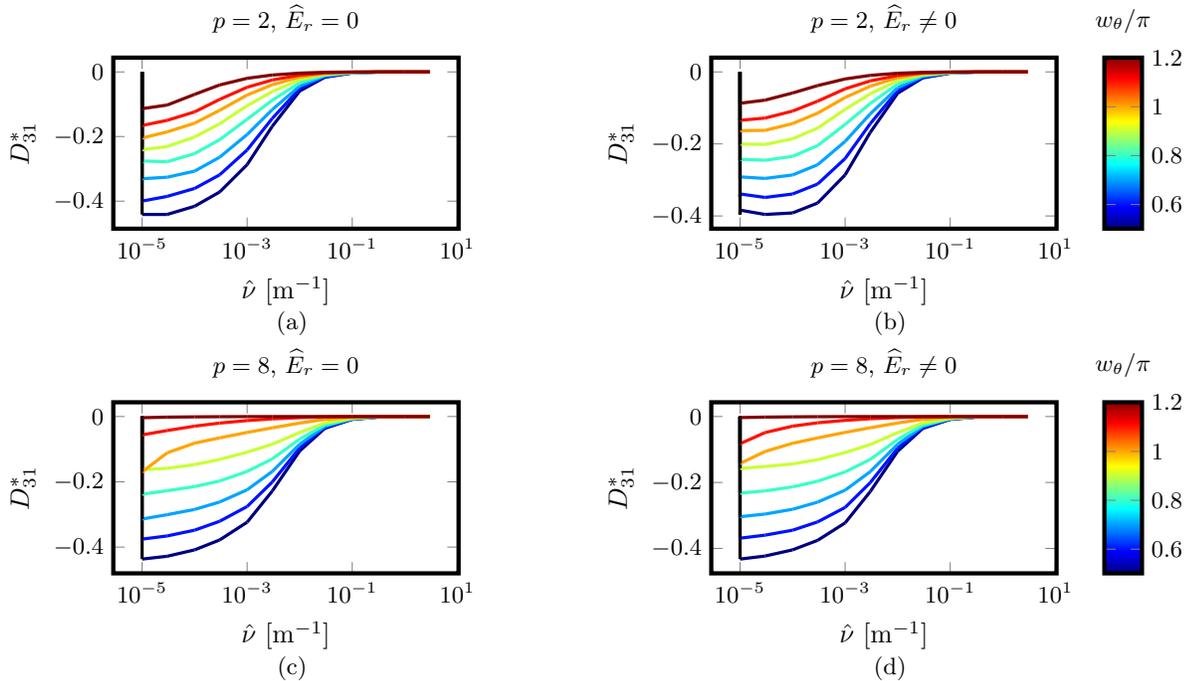
  
	\foreach \p in {4,16}
	{
		\tikzsetnextfilename{D31_vs_nu_pwQI_\p_Er_0} 
		\begin{subfigure}[t]{0.45\textwidth}
			\includegraphics{D31_vs_nu_pwQI_\p_Er_0} 
%
%
			\caption{}
			\label{subfig:D31_vs_nu_pwQI_\p_Er_0}
		\end{subfigure}
		\tikzsetnextfilename{D31_vs_nu_pwQI_\p_Er_1e-3} 
		\begin{subfigure}[t]{0.45\textwidth}
			\includegraphics{D31_vs_nu_pwQI_\p_Er_1e-3}
%
%
			\caption{}
			\label{subfig:D31_vs_nu_pwQI_\p_Er_1e-3}
		\end{subfigure}
		
	}	 
	\caption{Bootstrap current coefficient $D_{31}^*$ as a function of $\hat{\nu}$ and $w_\theta$ for $\widehat{E}_r \ne 0$}
	\label{fig:Dij_vs_nu_pwQI_Er_1e-3}
\end{figure*}
\begin{figure*}
	\centering	%
	\begin{subfigure}[t]{0.45\textwidth}%
		\tikzsetnextfilename{D31_vs_pi_factor_nu_1e-5_Er_0}
		\includegraphics{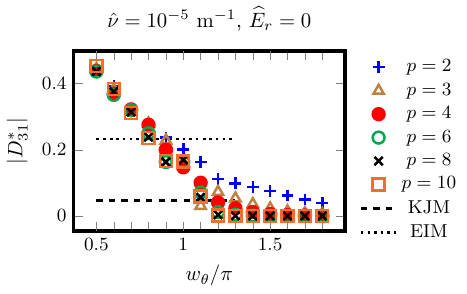}
		\caption{}
		\label{subfig:D31_vs_pi_factor_nu_1e-5}
	\end{subfigure} 
	\begin{subfigure}[t]{0.45\textwidth}		
		\tikzsetnextfilename{D31_vs_pi_factor_nu_1e-5_Er_1e-3}		
		\includegraphics{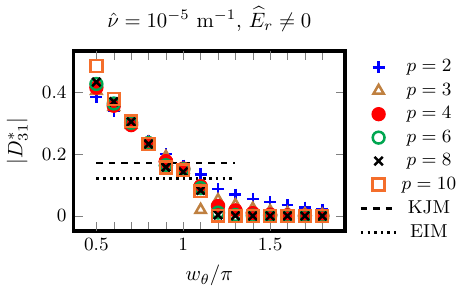}
		\caption{}
		\label{subfig:D31_vs_pi_factor_nu_1e-5_Er_1e-3}
	\end{subfigure} 
	\caption{Bootstrap current coefficient $D_{31}^*$ as a function of $w_\theta$ and $p$ for $\hat{\nu}=\num{e-5}$ $\text{m}^{-1}$.}
\end{figure*}

We can use the approximately pwO fields represented in figure \ref{fig:Magnetic_field_strength_pwO_QI} to precise our previous comment about how $\iota$ guarantees that $J$ is a flux-function piecewisely in the limit $p\rightarrow\infty$. When $\iota$ is given by (\ref{eq:iota_pwO}), two field lines connect the four corners of the parallelogram. These field lines are indicated in figures \ref{subfig:pwOMagneticField_pow_20_wa0.5_pi} and \ref{subfig:pwOMagneticField_pow_20_wa0.9_pi} with a white dashed line and their intersections with the parallelogram define several regions in the $(\theta,\zeta)$ plane. As for each region all bounce points lie on two parallel segments of the parallelogram, the angular distance between bounce points does not depend on the field line chosen. Besides, as for each region $B$ is also constant, then $J$ must also be the same for any field line belonging to that region. Thus, the orbit-averaged drift that trapped particles experience at each region is zero. Across the two field lines that delimit different regions, the value of $J$ can change abruptly. For an exactly pwO field, these transitions do not contribute to radial transport in the $1/\nu$ regime (see \cite{velasco2024piecewise}). For our model, the benignancy of transitioning particles is guaranteed by the fact that in the limit $p\rightarrow\infty$ the isolines have pointy corners. Hence, in the limit $p\rightarrow \infty$ any field where $B$ and $\iota$ are appropriately defined by (\ref{eq:Exponential_pwO}) and (\ref{eq:iota_pwO}) (e.g. as in the manner explained in \ref{sec:Appendix_pwO_B}) would have $\epseff=0$. From figures \ref{subfig:pwOMagneticField_pow_20_wa1.0_pi} and \ref{subfig:pwOMagneticField_pow_20_wa1.3_pi}, we can verify that increasing $w_\theta$ beyond $\pi$ forces the isolines of $B$ to close poloidally. As a consequence, at some point in the scan, the different classes of trapped particles disappear and $J$ becomes constant on the whole flux-surface, making the resulting field quasi-poloidally symmetric (a particular case of QI). Note from figure \ref{subfig:pwOMagneticField_pow_20_wa1.0_pi} that for $w_\theta=\pi$ the isoline $B=\Bmin$ is not poloidally closed due to the finiteness of $p$. In the limit $p\rightarrow\infty$, this isoline would close at precisely $w_\theta=\pi$. We recall that, by definition, the integer power $p$ represents the proximity to piecewise omnigenity of the model field. Similarly, $w_\theta$ controls closeness to quasi-isodynamicity of the configuration.

A very attractive feature of pwO fields is that rough approximations to an exactly pwO field can have low levels of radial neoclassical transport \cite{velasco2024piecewise}. In particular, we will see that using the model given by equations (\ref{eq:Exponential_pwO}) and (\ref{eq:iota_pwO}) for $p=2$ (the lowest value of $p$ considered), a banana-like regime \cite{Landreman_PreciseQS} appears between the plateau and the deleterious $1/\nu$ regime. Thus, the reduction of radial neoclassical transport appears for values of $p$ for which the magnetic field varies in a scale compatible with rigorous neoclassical theory \cite{velasco2024piecewise}. For this reason, pwO magnetic fields are very promising as an ideal design goal for optimization. Therefore, in this section, we will explore the parameter space $(p,w_\theta)$ to identify portions of it which have simultaneously small levels of radial and parallel transport. In the light of what has been exposed we expect $\Dij{11}$ to decrease with increasing $p$ and, for each fixed $p$, $|\Dij{31}|$ to be a monotonically decreasing function of $w_\theta$. Besides, for sufficiently large $w_\theta$ we expect $|\Dij{31}|$ to be a monotonically decreasing function of $p$.

In order to verify numerically our theoretical expectations, we have computed the monoenergetic coefficients $\Dij{11}$ and $\Dij{31}$ for collisionalities $\hat{\nu}\in[\num{e-5},3]$ $\text{m}^{-1}$ and radial electric field $\widehat{E}_r\in\{0,\num{e-3}\}$ $\text{kV}\cdot\text{s} /\text{m}^{2}$. This scan in collisionality and radial electric field has been carried out for approximately pwO fields constructed as indicated in \ref{sec:Appendix_pwO_B} for $p\in[2,10]$ and $w_\theta/\pi \in[0.5,1.9]$. In figure \ref{fig:Dij_vs_nu_pwQI_Er_0} the result of the scan in collisionality is shown for $\widehat{E}_r=0$ for $p=2$ and $p=8$. In colours, the value of $w_\theta/\pi$ for each case is displayed. As it was mentioned, we can see from the curve of $D_{11}^*$ plotted in figure \ref{subfig:D11_vs_nu_pwQI_4_Er_0} that even for $p=2$ a banana-like regime appears for $w_\theta/\pi \le 0.8 $. For higher values of $p$ the situation is similar, as shown in \ref{subfig:D11_vs_nu_pwQI_16_Er_0} for $p=8$. For $\hat{\nu}\gtrsim \num{e-4}$, increasing $w_\theta$ diminishes the value of $D_{11}^*$ for all values of $p$ considered. This happens even when $w_\theta$ is increased beyond $ 0.8\pi$ and the banana-like regime disappears. For large values of $w_\theta$ the width in $\hat{\nu}$ of the plateau region seems to increase. The apparent spread of the plateau region is due to the reduction of the $D^*_{11}$ in the low collisionality region where the $1/\nu$ regime would appear in a magnetic field far from omnigenity. The reduction of the value of $D_{11}^*$ in the plateau region with $w_\theta$ is due to the fact that for high values of $w_\theta$, the magnetic field becomes almost constant along $\theta$ \cite{EmiliaRodriguez1987}. For $\hat{\nu}\lesssim \num{e-4}$ there is, however, an increase of transport in the $1/\nu$ regime for $w_\theta\sim\pi$ as figure \ref{subfig:D11_vs_nu_pwQI_16_Er_0} reveals. This effect is seen in more detail in \ref{subfig:D11_vs_pi_factor_nu_1e-5} where the value of $D_{11}^*$ for $\hat{\nu}=\num{e-5}$ and $\widehat{E}_r=0$ is shown. We can see from this figure that $D_{11}^*(\hat{\nu}=\num{e-5})$ grows as it approaches $w_\theta=\pi$ and that this growth is ameliorated when $p$ is increased. The peak of radial transport when $w_\theta$ grows ($w_\theta<\pi$) may be caused by a combination of the finiteness of $p$ and the fact that orbits become shorter in the region between the tilted sides of the parallelogram defined by (\ref{eq:pwO_tilted_sides}) as $w_\theta$ grows. Due to the smaller value of $J$ in this narrow region, minimizing $\partial_{\alpha}{J}/J$ for these orbits requires a larger value of $p$ when $w_\theta$ grows. Nevertheless, this increase in radial transport is not larger than a factor of 2 for any of the cases considered. When there is a radial electric field, increasing $w_\theta$ also produces a flattening of the $D_{11}^*$ curve from plateau to low collisionality. Again, there is an increasing of the radial transport coefficient at low collisionality when $w_\theta\sim\pi$. This effect can be observed in more detail in figure \ref{subfig:D11_vs_pi_factor_nu_1e-5_Er_1e-3}. As for the $1/\nu$ regime, the growth in $D_{11}^*$ is less pronounced for the largest value of $p$ considered.

\captionsetup[sub]{skip=-1.75pt, margin=40pt}
\begin{figure}
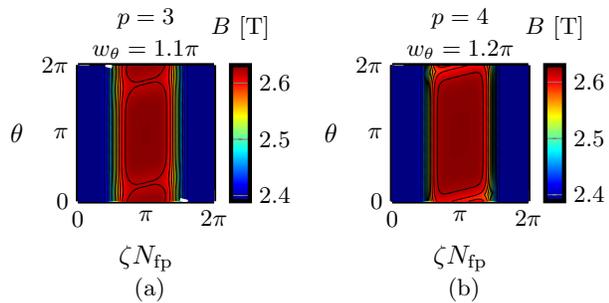

	\centering
	\foreach \w in {1.1}
	{ 
		\foreach \p in {6}
		{%
			\tikzsetnextfilename{pwQI_B_\w_\p_small_D31}%
			\begin{subfigure}[t]{0.23\textwidth} 
			   \includegraphics{pwQI_B_\w_\p_small_D31}%
			   \caption{}
			   \label{subfig:pwOMagneticField_Isolines_pow_\p_wa\w_pi}
	    	\end{subfigure}%
		}%
	} 
	\foreach \w in {1.2}
	{ 
		\foreach \p in {8}
		{%
			\begin{subfigure}[t]{0.23\textwidth} 
				\includegraphics{pwQI_B_\w_\p_small_D31}%
				\caption{}
				\label{subfig:pwOMagneticField_Isolines_pow_\p_wa\w_pi}
			\end{subfigure}%
		}%
	} 
	\caption{Magnetic field strength $B$ of those approximately pwO magnetic fields with smaller $D_{31}^*$ ($w_\theta=1.2\pi$) than the KJM configuration.}
	\label{fig:Magnetic_field_strength_pwO_QI_small_D31}
\end{figure}

In regard to the bootstrap current coefficient, excepting a few cases of small $p$ at low collisionality, the results for finite and zero $\widehat{E}_r$ are very similar due to the extreme proximity to omnigenity of the magnetic fields considered. For small values of $p$, the effect of increasing $w_\theta$ is to reduce the value of $|D_{31}^*|$ uniformly, as shown in figures \ref{subfig:D31_vs_nu_pwQI_4_Er_0} and \ref{subfig:D31_vs_nu_pwQI_4_Er_1e-3}. For higher values of $p\ge 4$, in the region $w_\theta \sim \pi$, the $|D_{31}^*|$ curve changes its convexity in the range of collisionalities considered. Thus, at the lowest collisionality, the value of $D_{31}^*$ for $w_\theta=\pi$ (orange curve) can be approximately equal to the one for $w_\theta=0.9\pi$ (lime curve). Nevertheless, as can be seen in figures \ref{subfig:D31_vs_pi_factor_nu_1e-5} and \ref{subfig:D31_vs_pi_factor_nu_1e-5_Er_1e-3}, the effect of increasing $w_\theta$ is to reduce $|D_{31}^*|$ at low collisionality. As expected, we can observe from comparing figures \ref{subfig:D31_vs_nu_pwQI_4_Er_0} and \ref{subfig:D31_vs_nu_pwQI_16_Er_0} or \ref{subfig:D31_vs_nu_pwQI_4_Er_1e-3} and \ref{subfig:D31_vs_nu_pwQI_16_Er_1e-3} that, for fixed $w_\theta>\pi$, the reduction in $|D_{31}^*|$ is  typically more pronounced for bigger values of $p$. We can see from figure \ref{subfig:D31_vs_pi_factor_nu_1e-5} that when $\widehat{E}_r=0$ and $w_\theta/\pi \ge 1.2$ and $p\ge4$, the value of $|D_{31}^*|$ is smaller than that of the KJM configuration (also without $\widehat{E}_r$). Another case with such small value of the bootstrap current coefficient is the case $p=3$ and $w_\theta/\pi=1.1$. These results suggest that it is possible to design stellarators that deviate from QI to approach pwO with small levels of both radial and parallel transport. However, in this first exploration, the results indicate that it is necessary to be close to QI to have small $|D_{31}^*|$. In order to have a $|D_{31}^*|$ value equal or smaller than that of the KJM configuration without $\widehat{E}_r$ we need at least $p=3$ and $w_\theta/\pi=1.1$ or $p=4$ and $w_\theta/\pi=1.2 $. By inspecting figures \ref{subfig:pwOMagneticField_Isolines_pow_6_wa1.1_pi} and \ref{subfig:pwOMagneticField_Isolines_pow_8_wa1.2_pi}, we can check that most of the isolines of $B$ for this case are poloidally closed. Interestingly, we can see from this figure that those isolines which do not close poloidally are located around $\Bmax$. This is in agreement with what the numerical results shown in section \ref{sec:Correlations} suggest. In section \ref{sec:Correlations} we verified that minimizing the proxy $\sigma^2(B(\theta,0))$ (equivalently $\sigma^2(\Bmax^{\text{r}})$) did not entail a reduction of $|D^*_{31}|$. We want to stress out that this exploration is far from being exhaustive and the configuration space of pwO fields has to be investigated further in future work. There are many other ways to transition from pwO to QI which might allow larger deviations from QI without compromising the small levels of radial and parallel neoclassical transport. For instance, one could consider a QI magnetic field as the one shown in \ref{subfig:pwOMagneticField_pow_20_wa1.3_pi} and use $\widetilde{B}$ as defined in (\ref{eq:Exponential_pwO}) to assess the neoclassical impact of a small pwO perturbation. It seems reasonable to expect that allowing some of the isolines of $B$ of a nearly QI field to close as in a pwO field, will allow to relax the stringent condition that being exactly QI imposes without increasing significantly the bootstrap current nor radial transport.

Finally, we point out that an exactly pwO magnetic field cannot be represented with a Fourier series without suffering the Gibbs phenomenon (further details in \ref{sec:Appendix_pwO_B}). This phenomenon is caused by the discontinuity of $B$ in the perimeter of the parallelogram. Even though we have considered only finite values of $p$, due to the large gradients of $B$ in the vicinity of the perimeter, approximating $\widetilde{B}$ with a Fourier series requires a large amount of modes $\{B_{mn}\}$ with big mode numbers $(m,n)$. This unusually broad spectrum of $B$ for high $p$ implies that, in order to solve the drift-kinetic equation at low collisionality, the spatial and Legendre resolutions must be very large. For instance, for $p= 10$ and $\hat{\nu}=\num{e-5}$, calculating the monoenergetic coefficients using {\MONKES} required around $ 12000 $ discrete Fourier modes and $200$ Legendre modes. For this extremely (and unusually) large spatial resolution, the wall-clock time for computing the monoenergetic coefficients for each pair $(\hat{\nu},\widehat{E}_r)$ was of approximately 14 minutes while running using 30 cores of CIEMAT's cluster XULA. Hence, the investigation of pwO magnetic fields would have been much more difficult (if not practically impossible) without a fast neoclassical code like {\MONKES}.

\endgroup
\section{Conclusions}
In this work we have used the new neoclassical code {\MONKES} for evaluating a large number of magnetic configurations which are close to be QI. In the first part of the paper we have
 demonstrated that, although relatively effective, the indirect approach to neoclassical optimization is not very efficient. In particular, it is specially inefficient for reducing parallel transport. This result stresses out the importance of optimizing directly the bootstrap current in the next generation of optimized stellarators. Fortunately, thanks to {\MONKES} speed of computation, this optimization strategy can be realized. 
 
 In the second part of this paper, a detailed study of neoclassical transport in nearly pwO magnetic fields has been carried out. Through this evaluation we have verified that pwO fields can, in principle, have simultaneously small radial transport and bootstrap current by approaching quasi-isodynamicity. Thus, the results suggest that the concept of piecewise omnigenity may be used for shaping the isolines of $B$ in order to have a mixture of a QI field with an approximately pwO magnetic field with small levels of radial and parallel neoclassical transport. Hopefully, this new ideal design goal will facilitate better trade-offs to meet reactor-relevant criteria. Given the numerical results, it seems possible that, through direct optimization using {\MONKES}, the different trade-offs between competing proxies naturally produce deviations from quasi-isodynamicity towards piecewise omnigenity without compromising the smallness of the bootstrap current.

 \section*{Acknowledgements}
 This work has been carried out within the framework of the EUROfusion Consortium, funded by the European Union via the Euratom Research and Training Programme (Grant Agreement No 101052200 – EUROfusion). Views and opinions expressed are however those of the author(s) only and do not necessarily reflect those of the European Union or the European Commission. Neither the European Union nor the European Commission can be held responsible for them. This research was supported in part by grant PID2021-123175NB-I00, funded by Ministerio de Ciencia, Innovación y Universidades/Agencia Estatal de Investigación/10.13039/501100011033 and by ERDF/EU.


\appendix
\section{Normalization of the monoenergetic coefficients}\label{sec:Appendix_monoenergetic_normalization}
The monoenergetic coefficients $\Dij{11}$ and $\Dij{31}$ defined in \cite{escoto2024monkes}, are related to their normalized versions $D_{11}^*$ and $D_{31}^*$ defined in \cite{Beidler_2011} as
\begin{align}
	D_{11}^* & = \frac{ 8R  B_0^2 \iota}{\pi} K_{11} \Dij{11}, 
	\\
	D_{31}^* & =  \iota B_0 \sqrt{\frac{a}{R}} K_{31} \Dij{31}. 
\end{align}
Here, $R$ and $a$ are, respectively, the major and minor radius of the device. $B_0$ is a reference value for $B$ on the flux-surface and $\iota$ is the rotational transform. The normalization factors $K_{31}=\dv*{r}{\psi}$, $K_{11}=K_{31}^2$ change from the radial coordinate $\psi$ to $r=a\sqrt{\psi/\psi_{\text{lcfs}}}$ where $2\pi\psi$ is the toroidal flux of $\vb*{B}$ enclosed by the flux-surface and $\psi_{\text{lcfs}}$ is the label of the last closed flux-surface. 

\section{Equivalency between $\sigma^2(B(\theta,0))$ and $\sigma^2(\Bmax)$ for the optimization campaign}\label{sec:Appendix_VBM}
\begin{figure*} 		 
	\tikzsetnextfilename{D31_vs_D11_KN_VBM_0100_Er_0}	
	\begin{subfigure}[t]{0.45\textwidth}	 		
        \includegraphics{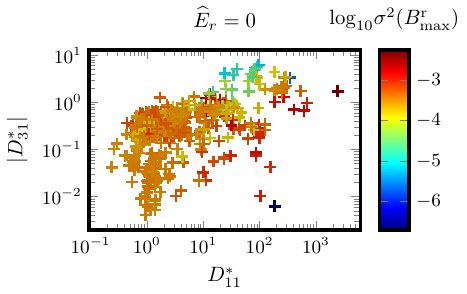}	
		\caption{}
		\label{subfig:D31_vs_D11_KN_VBM_0100_Er_0}
	\end{subfigure} 
	\tikzsetnextfilename{D31_vs_D11_KN_VBM_0100_Er_1e-3}	
	\begin{subfigure}[t]{0.45\textwidth}	 		
		\includegraphics{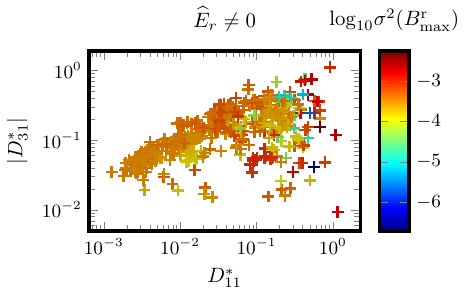}	
		\caption{}
		\label{subfig:D31_vs_D11_KN_VBM_0100_Er_1e-3}
	\end{subfigure} 

	\tikzsetnextfilename{D31_vs_KN_VBM_0100_Er_0}	
	\begin{subfigure}[t]{0.45\textwidth}	 		
        \includegraphics{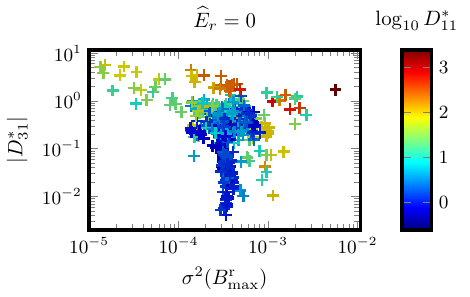}	
		\caption{}
		\label{subfig:D31_vs_KN_VBM_0100_Er_0}
	\end{subfigure} 
	\tikzsetnextfilename{D31_vs_KN_VBM_0100_Er_1e-3}	
	\begin{subfigure}[t]{0.45\textwidth}	 		
		\includegraphics{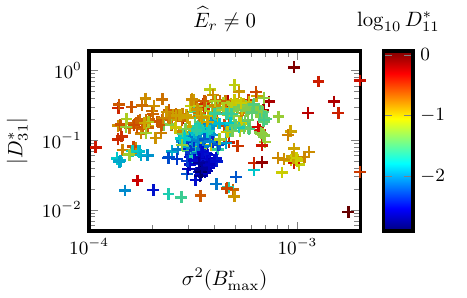}
		\caption{}
		\label{subfig:D31_vs_KN_VBM_0100_Er_1e-3}
	\end{subfigure} 

	\tikzsetnextfilename{D31_vs_KN_VBM_0100_Er_0_restricted}	
	\begin{subfigure}[t]{0.45\textwidth}	 		
		\includegraphics{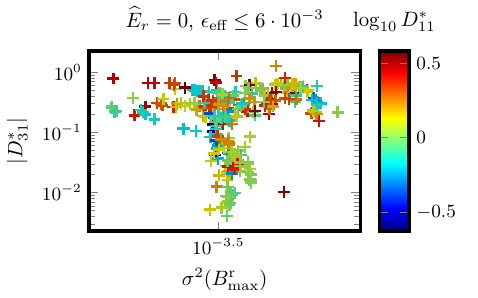}	
		\caption{}
		\label{subfig:D31_vs_KN_VBM_0100_Er_0_restricted}
	\end{subfigure} 
	\tikzsetnextfilename{D31_vs_KN_VBM_0100_Er_1e-3_restricted}	
	\begin{subfigure}[t]{0.45\textwidth}	 		
		\includegraphics{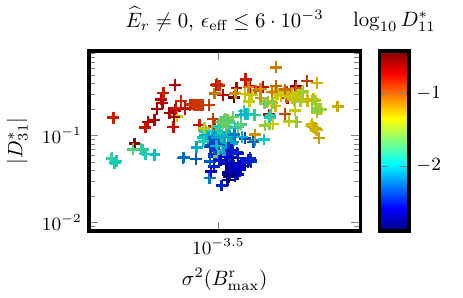}	
		\caption{}
		\label{subfig:D31_vs_KN_VBM_0100_Er_1e-3_restricted}
	\end{subfigure}

	\caption{Relation of the radial transport $D_{11}^*$ and bootstrap current $D_{31}^*$ coefficients with $\sigma^2(\Bmax^{\text{r}})$.}
	\label{fig:Correlation_VBM}
\end{figure*}

In this appendix we illustrate the equivalency of the proxies $\sigma^2(B(\theta,0))$ and $\sigma^2(\Bmax)$ for the optimization campaign considered in section \ref{sec:Correlations}. For this we plot the results of the evaluation against $\sigma^2(\Bmax)$. By comparing figures \ref{subfig:D31_vs_D11_KN_VBM_0100_Er_0}-\ref{subfig:D31_vs_KN_VBM_0100_Er_1e-3_restricted}, respectively, with \ref{subfig:D31_vs_D11_KN_VB0_0100_Er_0}-\ref{subfig:D31_vs_KN_VB0_0100_Er_1e-3_restricted} we can see that both the distribution of points and the colour pattern are quite similar. Thus, the conclusions that were extracted for $\sigma^2(B(\theta,0))$ in section \ref{sec:Correlations} are applicable to $\sigma^2(\Bmax)$.

\section{Construction of the approximately pwO field using equations (\ref{eq:Exponential_pwO}) and (\ref{eq:iota_pwO})}\label{sec:Appendix_pwO_B}
In this appendix we will explain how the approximately pwO magnetic fields from section \ref{sec:pwO_QI} are constructed using (\ref{eq:Exponential_pwO}) and (\ref{eq:iota_pwO}). In section \ref{sec:pwO_QI} we mentioned that $\widetilde{B}$, as defined in (\ref{eq:Exponential_pwO}), can only define an exactly pwO field for $w_\theta\le \pi - |t_\zeta|w_\zeta$. Note that the inadequacy of $\widetilde{B}$ for $w_\theta> \pi -|t_\zeta|w_\zeta$ is inherited from the inadequacy of the function 
  \begin{align}
		\eta(\theta,\zeta)
		& := 
		\exp(
		- 
		\left(
		\frac{\theta-\theta_{\text{c}} - t_\zeta\left(\zeta-\zeta_{\text{c}}\right)}{w_\theta}
		\right)^{2p} 
		)
		\nonumber
		\\
		&\times
		\exp(
		-
		\left(
		\frac{\zeta-\zeta_{\text{c}}}{w_\zeta}
		\right)^{2p} 
		),
  \end{align}
  in this same region of the parameter space. 
  
  We can circumvent this problem by defining the following auxiliary functions
  \begin{align}
  	\eta_k(\theta,\zeta) : = \eta(\theta + 2k\pi,\zeta)
  \end{align}
  for $k\in\mathbb{Z}$, 
  \begin{align}
  	\eta^{\text{s}} : = 
  	\sum_{k=-1}^{1}
  	\eta_k
  	, 
  \end{align}
  \begin{align}
  	\eta^{\text{m}} : = 
  	\max_{k\in\{-1,0,1\}}  	
  	\{\eta_k\},
  \end{align}
  and 
  \begin{align}
  	\eta^{H} : = 
  	H
  	\left(
  	\eta^{\text{s}} -\Bmax
  	\right)
  	\eta^{\text{m}} 
  	+
  	H
  	\left(
  	\Bmax - \eta^{\text{s}} 
  	\right)
  	\eta^{\text{s}}
  \end{align}
  where $H(x)=1$ for $x\ge0$ and $H(x)=0$ otherwise. Note that $\eta^{\text{s}}$, $\eta^{\text{m}}$ and $\eta^H$ are equal in the limit $p\rightarrow\infty$ for all $w_\theta\le\pi$. In this limit we could define the magnetic field strength of an exactly pwO for $w_\theta > \pi - |t_\zeta| w_\zeta$ using e.g. $\eta^{\text{s}}$ as $B=\Bmin + (\Bmax-\Bmin)\eta^{\text{s}}$. 
  
  For finite $p$, $\eta^{\text{s}}$, $\eta^{\text{m}}$ and $\eta^H$ are different and we will need to choose which one use at each point of the $(\theta,\zeta)$ plane for different values of $w_\theta$. In addition, $w_\theta$ must be defined for values greater than $\pi$ (recall that for finite $p$ the isoline of $\Bmin$ does not close poloidally at $w_\theta=\pi$). For $w_\theta > \pi$, $\eta^{\text{m}}$ is not differentiable and $\eta^{\text{s}}$ can be larger than 1 at some points. On the other hand, the function $\eta^H$ is always smaller or equal than 1 at the expense of not being differentiable at a few points. Using these functions we can define 
  \begin{align}
  	\eta_{\text{pwO}} = 
  	\begin{dcases}
  		\eta^{\text{m}}, & w_\theta < \pi,
  		\\
  		 \eta^{\text{s}}, & w_\theta = \pi,
  		\\
  		\eta^{H}, & w_\theta > \pi,
  	\end{dcases}  	
  \end{align}
  and 
  \begin{align}
  	\BpwO = 
  	\Bmin 
  	+ 
  	(\Bmax-\Bmin)
  	\eta_{\text{pwO}}.
  \end{align}
  
  Finally, we approximate $\BpwO$ using a stellarator-symmetric Fourier series
  \begin{align}
  	B = \sum_{m,n} B_{mn}\cos(m\theta + n\Nfp \zeta),
  	\label{eq:Magnetic_field_strength_approximate_pwO}
  \end{align}
  where $ B_{mn}$ are the discrete Fourier modes of $\BpwO$. Hence, if we fix the parameters $\{\theta_{\text{c}},\zeta_{\text{c}},w_\theta,w_\zeta,t_\zeta\}$ that define $\eta$ and, additionally, $\Bmin$ and $\Bmax$, we can define the magnetic field strength $B$ using (\ref{eq:Magnetic_field_strength_approximate_pwO}) and $\iota$ using (\ref{eq:iota_pwO}) for each pair $(p,w_\theta)$. It is important to remark that, in the limit $p\rightarrow\infty$, representation (\ref{eq:Magnetic_field_strength_approximate_pwO}) will suffer from the Gibbs phenomenon due to the discontinuity at the perimeter of the parallelogram (where $\BpwO$ abruptly changes from $\BpwO=\Bmin$ to $\BpwO=\Bmax$). For finite $p$, Gibbs phenomenon does not appear around the parallelogram, but as $\BpwO$ still changes in a very short length scale, the modes $B_{mn}$ will have very large mode numbers $m$ and $n$ making the Fourier spectra extremely broad (in comparison with standard stellarator configurations). Finally we nuance that, for finite $p$, Gibbs phenomenon does appear as $\BpwO$ is not periodic in $\theta$ nor $\zeta$. However, this form of Gibbs phenomenon is benign. As the exponential $\eta_{\text{pwO}}$ is not periodic we can find values of $\zeta$ and $\theta$ where $\BpwO(0,\zeta)\ne\BpwO(2\pi,\zeta)$ and $\BpwO(\theta,0)\ne\BpwO(\theta,2\pi/\Nfp)$ respectively. However, due to the attenuation produced by the exponential, the differences $\BpwO(0,\zeta)-\BpwO(2\pi,\zeta)$ and $\BpwO(\theta,0)-\BpwO(\theta,2\pi/\Nfp)$ are of the order of the round-off error and have no significant impact on the modes $B_{mn}$.
  
  The parameters required for defining $\eta$ have been selected so that the magnetic configuration resembles that of Wendelstein 7-X KJM at $s=0.2$. For each pair $(p,w_\theta)$, the values of $\Bmax$ and $\Bmin$ are selected so that the discrete Fourier mode $B_{00}$ of $B$ matches that of the KJM configuration (however $\Bmax$ and $\Bmin$ vary very little between fields). This Fourier mode is also the reference value of $B$ on the flux-surface, which we denote by $B_0$. In table \ref{tab:pwO_parameters} the values of the remaining parameters that define $\eta$ for each pair $(w_\theta,p)$ are shown. Note that $\theta_{\text{c}}$ and $\zeta_{\text{c}}$ are selected so that $\eta$ (and therefore $B$) satisfies stellarator-symmetry. Also note that fixing $t_\zeta$ and $w_\zeta \Nfp$ also determines $\iota=-t_\zeta$ via constraint (\ref{eq:iota_pwO}). 
  
  In order to compute the monoenergetic coefficients $\Dij{ij}$ we also need to specify $\{\Nfp,B_\theta,B_\zeta\}$ where $B_\theta$ and $B_\zeta$ are the covariant components of $\vb*{B}$ in Boozer coordinates. In addition, for computing their normalized versions $D_{ij}^*$ we need to specify the minor radius $a$ and major radius $R$ along with the radial derivative of the toroidal flux (divided by $2\pi$) $\dv*{\psi}{r}$. The quantities $\dv*{\psi}{r}$, $B_\theta$ and $B_\zeta$ are those of Wendelstein 7-X KJM at $s=0.2$. In table \ref{tab:pwO_parameters_KJM}, the remaining parameters required to compute the normalized monoenergetic coefficients $D_{ij}^*$ are listed. The minor and major radius are approximated employing, respectively, estimates $a_{\text{lar}}$ and $R_{\text{lar}}$, which are valid for a large aspect ratio stellarator 
  \begin{align}
  	a_{\text{lar}} & :=  \frac{ \left| \dv*{\psi}{r} \right|}{B_0} , 
  	\\
  	R_{\text{lar}} & :=  \frac{ \left|B_\zeta \right|}{B_0 }.
  \end{align}

\begingroup 
\captionsetup[sub]{skip=-1.75pt, margin=40pt}
\newcommand{\LABEL}[2]{\label{\JointString{#1}{#2}}}
\begin{figure*}
	\centering
	\foreach \p in {4, 8, 20}
	{ 
		\foreach \w in {0.9, 1.0, 1.1, 1.2}
		{%
			\tikzsetnextfilename{Appendix_pwQI_B_\w_\p}
			\begin{subfigure}[t]{0.24\textwidth}
				\includegraphics{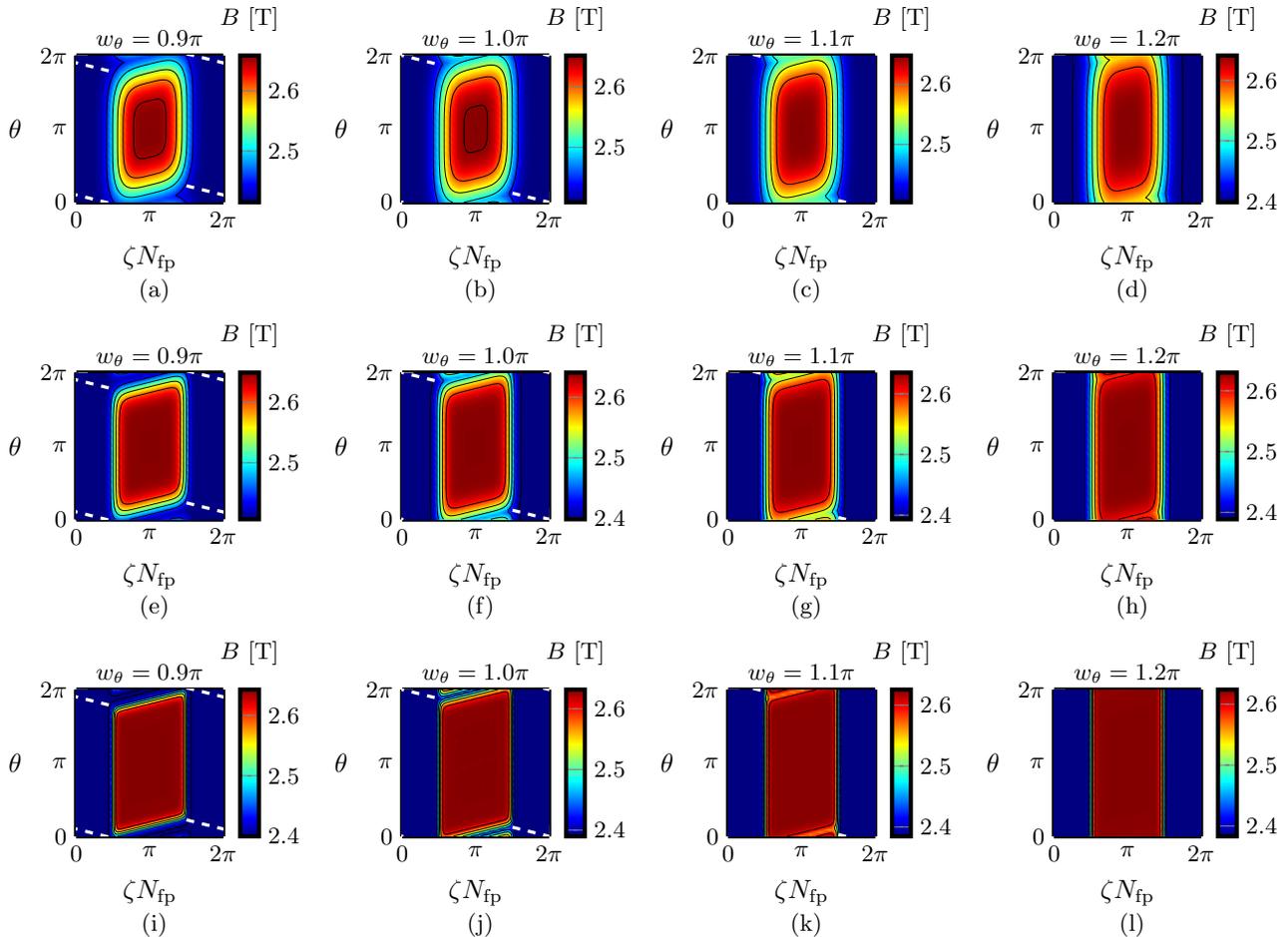}
				\caption{}
				\label{subfig:Appendix_pwOMagneticField_pow_\p_wa\w_pi} 
			\end{subfigure}
			%
		}  
		
	} 
	\caption{Magnetic field strength $B$ for the parameter scan in pwO configuration space. $p=2$ (top row), $p=4$ (middle row) and $p=10$ (bottom row).}
	\label{fig:Appendix_Magnetic_field_strength_pwO_QI}
\end{figure*}
\endgroup

\begin{table}[]
	\centering
	\begin{tabular}{@{}ccccc@{}}
		\toprule
		$\Nfp$ &    $w_\zeta \Nfp$ & $t_\zeta$    & $\theta_{\text{c}}$ & $\zeta_{\text{c}}$ \\ \midrule
		5&  $\pi/2$ & $1.242$ & $\pi$      & $\pi/N_{\text{fp}}$     \\ \bottomrule
	\end{tabular}
	\caption{Parameters selected for defining $\widetilde{B}$ . }
	\label{tab:pwO_parameters}
\end{table}

\begin{table}[]
	\centering
	\begin{tabular}{@{}cccccc@{}}
		\toprule
		$B_0$ & $\dv*{\psi}{r}$   & $B_\theta$ & $B_\zeta$ & $a_{\text{lar}}$ & $R_{\text{lar}}$\\ \midrule
		2.5003 &   0.5132   &  0    &   $-14.4$    & 0.205 & 5.76 \\ \bottomrule
	\end{tabular}
	\caption{Parameters that define the rescaling of the pwO magnetic field. $B_0$ is given in T, $\dv*{\psi}{r}$, $B_\theta$ and $B_\zeta$ are given in $\text{T}\cdot\text{m}$. The minor $a_{\text{lar}}$ and major radius $R_{\text{lar}}$ are given in m.}
	\label{tab:pwO_parameters_KJM}
\end{table}

In figure \ref{fig:Appendix_Magnetic_field_strength_pwO_QI}, the magnetic field strength $B$ is represented for $p\in\{2,5,10\}$ for the values $w_\theta/\pi\in\{0.9,1.0,1.1,1.2\}$, which are those of the transition from pwO to QI. The effect of increasing $p$ can be observed by looking the columns of figure \ref{fig:Appendix_Magnetic_field_strength_pwO_QI} from the top row ($p=2$) to the bottom row ($p=10$). As was mentioned in section \ref{sec:pwO_QI}, we can verify that increasing $p$ compresses the isolines between $\Bmin$ and $\Bmax$ and thus, the gradient of $B$ on the flux-surface is maximum in the surroundings of the perimeter of the parallelogram. The effect of increasing $w_\theta$ can be observed in figure \ref{fig:Appendix_Magnetic_field_strength_pwO_QI}, inspecting each row from the leftmost column ($w_\theta=0.9\pi$) to the rightmost one ($w_\theta=1.2\pi$). When $w_\theta<\pi$ the parallelogram fits in a single poloidal period. When $w_\theta$ is increased beyond $\pi$, the isolines of $B$ begin to close poloidally as expected. For $w_\theta \sim \pi$, we can see on figures \ref{subfig:Appendix_pwOMagneticField_pow_4_wa1.1_pi}, \ref{subfig:Appendix_pwOMagneticField_pow_4_wa1.2_pi}, \ref{subfig:Appendix_pwOMagneticField_pow_8_wa1.1_pi}, \ref{subfig:Appendix_pwOMagneticField_pow_8_wa1.2_pi}, \ref{subfig:Appendix_pwOMagneticField_pow_20_wa1.1_pi} and \ref{subfig:Appendix_pwOMagneticField_pow_20_wa1.2_pi} that the growth of the parallelogram with $w_\theta$ is periodic in the interval $\theta\in[0,2\pi]$. Thus, in the limit $w_\theta\rightarrow\infty$ (even for finite $p$), all isolines close poloidally and the magnetic field becomes quasi-poloidally symmetric. A particular case of quasi-poloidal symmetry with discontinuous $B$ can be attained if, for sufficiently large $w_\theta$, we take the limit $p\rightarrow\infty$. An approximation to this type of quasi-poloidal symmetry is shown in figure \ref{subfig:Appendix_pwOMagneticField_pow_20_wa1.2_pi}, consisting of a central poloidally closed \qmarks{stripe} of width $w_\zeta$ where $B \approx\Bmax$ and on the rest of the flux-surface $B\approx \Bmin$. Thus, as required, this scan permits to approach quasi-isodynamicity from pwO in a controlled manner by increasing $w_\theta$ and/or $p$.

\section*{References} 
\bibliographystyle{unsrt} 
\bibliography{refs}

\end{document}